\documentclass[12pt,oneside,final,a4paper]{article}

\title{$p$-adic discrete dynamical systems and their 
applications in physics and cognitive sciences}
\author{Andrei Khrennikov\footnote{Supported by the EU Human Potential Programme under 
contract N. HPRN-CT-2002-00279}\\
International Center for Mathematical\\
Modeling in Physics and Cognitive Sciences,\\
MSI, University of V\"axj\"o, S-35195, Sweden\\
Email:Andrei.Khrennikov@msi.vxu.se}

\begin{document}
\maketitle

\begin{abstract}
This review is devoted to dynamical systems in fields of $p$-adic numbers: origin of 
$p$-adic dynamics in 
$p$-adic theoretical physics (string theory, quantum mechanics and field theory, spin glasses),
continuous dynamical systems and discrete dynamical systems. The main attention is paid
to discrete dynamical systems - iterations of maps in the field of $p$-adic numbers (or their
algebraic extensions): 
ergodicity,  
behaviour of cycles, holomorphic dynamics. 
We also discuss applications of $p$-adic discrete dynamical systems
to cognitive sciences and psychology.
\end{abstract}

\section{Introduction}

The fields ${\bf Q}_p$ of $p$-adic numbers (where $p=2,3,\ldots, 1999, 
\ldots$ are prime numbers) were introduced by German mathematician K. Hensel 
at the end of 19th century, [1]. Hensel started with the following question: 

Is it possible to expend a rational number $x\in Q$ in a power series of the form
\begin{equation}
\label{S1}
x=\sum_{n=k}^\infty \alpha_np^n, \alpha_n=0, \ldots, p-1,
\end{equation}
where  $k=0, \pm 1, \pm 2, \ldots$ depends on $x$. Of course, this question was motivated by 
the existence of real expansions of rational numbers with respect to a $p$-scale:
\begin{equation}
\label{S2}
x=\sum_{n=-\infty}^k \alpha_np^n, \alpha_n=0, \ldots, p-1\;.
\end{equation}
Hensel knew, for example, that 
\[\frac{4}{3}=\sum_{n=-\infty}^0 2^{2n}.\]
He studied the possibility to expand $x=\frac{4}{3}$ in a series 
with respect to positive powers of $p=2:$
\[\frac{4}{3}=\sum_{n=0}^\infty \alpha_n2^n, \alpha_n=0,1.\]
Such rather innocent manipulations with rational numbers and series generated the idea that 
there exists some algebraic structure similar to the system of real numbers ${\bf R}$. 
K. Hensel observed that it is possible to
introduce algebraic operations (addition, subtraction, multiplication, division) on the set ${\bf Q}_p$ 
of all formal series
(\ref{S1}). Thus each ${\bf Q}_p$ has the structure of a number {\it{field}}. The field of rational numbers
${\bf Q}$ is a subfield of ${\bf Q}_p.$ In fact, 
the fields of $p$-adic numbers ${\bf Q}_p$ were first examples of infinite fields that 
differs from ${\bf Q}, {\bf R}$, 
${\bf C}$ and corresponding fields of rational functions.

It is possible to work in more general framework, 
namely to consider not only prime numbers $m$, but all natural numbers $p$ 
as bases of expansions. In principle, we can   do this. 
However, the corresponding number system is not in general a field.
If $m=p_1p_2,$ where $p_j$ are distinct prime numbers, then ${\bf Q}_m$ is not a 
field (there exists divisors of zero), but only a {\it{ring}},
i.e., division is not well defined. The field structure is very important to develop analysis. 
Therefore the main part of investigations is performed for prime $p.$ In particular, we consider only 
this case in this paper. 

The construction of new fields ${\bf Q}_p$ induced strong interest in number theory and algebra. Practically one
hundred years $p$-adic numbers were intensively used only in pure mathematics, mainly in number theory, see, 
for example, the classical book of Borevich and Schafarevich [2]. In particular, $p$-adic numbers 
are useful in investigations of some number-theoretical problems in the field of rational numbers ${\bf Q}.$ 
Typically if we can prove some fact for the field of real numbers ${\bf R}$ as well as for all fields
of $p$-adic numbers ${\bf Q}_p, p=2,3,\ldots,$ then we get the corresponding result for the field of rational 
numbers $Q,$ see [2] for the details.

The presence of the  structure of a topological field on ${\bf Q}_p$ gives the possibility to develop analysis for 
functions $f:{\bf Q}_p\rightarrow {\bf Q}_p.$ In particular, the derivative of such a 
function is defined in the usual way:
\[f^\prime(x)=\lim_{h\rightarrow 0}\frac{f(x+h)-f(x)}{h}, x, h\in {\bf Q}_p.\]

Of course, to perform limit-procedure, we need a topology on ${\bf Q}_p.$ It is  possible
to define on ${\bf Q}_p$ a $p$-adic
absolute value, {\it{valuation}}, $x\rightarrow |x|_p$ 
that has properties similar to the properties of the ordinary absolute value on ${\bf R}.$ 
The topology on the field 
${\bf Q}_p$ is defined by the metric $\rho_p(x,y)=|x-y|_p.$ The ${\bf Q}_p$ is a {\it{locally compact topological field}}
(i.e., the unit ball of ${\bf Q}_p$ is a compact set and all algebraic operations are continuous). 
The field of rational numbers ${\bf Q}$ is a dense subset of ${\bf Q}_p.$ Thus
${\bf Q}_p$ (as well as ${\bf R}) $is a completion of the field of rational numbers ${\bf Q}.$

 This review is devoted to applications of $p$-adic numbers in various
fields. However, we start with an extended historical review on $p$-adic (and more general non-Archimedean
or ulrametric) analysis and related fields. Such a review can be useful for researchers working in applications
of $p$-adic numbers.

The general notion of an absolute valued field was introduced by J. K\"urschak [3] in 1913 and a few years later
A. Ostrowski [4] described absolute values on some classes of fields, especially on the rationals.
In 1932 L. S. Pontryagin [5] proved that the only locally compact and connected topological division rings are the
classical division rings and N. Jacobson [6]  began the systematic study of the structure of locally compact rings.
We should also mention the theory of  Krull valuations [7] and the paper by S. MacLane [8] 
that began the study of valuations on polynomial rings. We recall that I. Kaplansky initiated
the general theory of topological rings, see e.g. [9].

In 1943 I. R. Shafarevich [10] found  necessary and sufficient
conditions for a topological field to admit an absolute value
preserving the topology and D. Zelinsky [11]  
characterized the topological fields admitting a non-Archimedean valuation.
We also mention the work [12] of H.-J. Kowalsky, who described locally compact fields.

First fundamental investigations in $p$-adic analysis were done by K. Mahler [13] 
(differential calculus, differential and difference equations), M. Krasner 
[14], [15] (topology on ${\bf Q}_p$, the notion of an ultrametric space 
\footnote{One of the important features of the $p$-adic metric $\rho_p$ is that it is an ultrametric. 
It satisfies to the strong triangle inequality, see section 2. Metric spaces in that the
strong triangle inequality holds true (ultrametric or non-Archimedean spaces) were actively used in analysis,
since Krasner's work [14]. In topology these spaces were actively studied, since the works of F. Hausdorff [16].},
$p$-adic analytic functions), E. Motzkin and Ph. Robba [17], [18] (analytic functions),
Y. Amice [19] (analytic functions, interpolation, Fourier analysis), M. Lazard [20] (zeros of 
analytic functions), A. Monna [21] (topology, integration), 
T. A. Springer [22] (integration, theory of $p$-adic Hilbert spaces).

In the period between 1960 and 1987 $p$-adic analysis developed intensively due to pure mathematical 
self-motivations. Main results of these investigations are collected in the nice book of W. Schikhof
[23] that is really a fundamental encyclopedia on $p$-adic analysis, recently there was published 
another excellent book on $p$-adic analysis -- namely the book [24] of A. M. Robert, see also 
the book [25] of A. Escassut on $p$-adic analytic functions and the book [26] of P.-C. Hu and
C.-C. Yang on non-Archimedean theory of meromorphic functions. An important approach 
to the non-Archimedean analysis is based on the theory of {\it rigid analytic spaces},
see  J. T. Tate [27], S. Bosch, L. Gerritzen, H. Grauert, U. G\"untzer, Ya. Morita, R. Remmert, 
[28]--[37]. We also mention works of Kubota, Leopoldt, Iwasawa, Morita [38]-[43] (ivestigations
on $p$-adic $L$ and $\Gamma$-functions).

This $p$-adic analysis is analysis for maps $f: {\bf Q}_p \to {\bf Q}_p$ (or finite or infinite algebraic extensions
of ${\bf Q}_p).$ And the present paper is devoted to the theory of dynamical systems based on such maps.

However, not only maps $f: {\bf Q}_p \to {\bf Q}_p,$ but also maps
maps $f: {\bf Q}_p \to {\bf C}$  are actively used in $p$-adic mathematical physics. 
Analysis for complex valued functions of the $p$-adic variable was intensively developed 
in general framework of the Fourier analysis on locally compact groups. 
There was developed a theory of distributions on locally compact disconnected fields (and, in particular,
fields of $p$-adic numbers). There were obtained fundamental results on theory of non-Archimedean
representations. Fundamental investigations
in this domain were performed in early 60th by I. M. Gelfand, M. I. Graev and I. I.  Pjatetskii-Shapiro
[44]-[48] (see also papers of M. I. Graev and R. I. Prohorova [49], [50],
A. A. Kirillov and R. R. Sundcheleev [51] and A. D. Gvishiani [52], P. M. Gudivok [53], A. V. Zelevinskii [54],
A. V. Trusov [55], [56]).

The first (at least known to me) publication on the possibility to use the $p$-adic space-time
in physics was article [57] of  A. Monna  and  F. van der Blij. Then E. Beltrametti  and G. Cassinelli
tried to use $p$-adic numbers in quantum logic [58]. But they obtained the negative result: $p$-adic numbers
could not be used in quantum logic. The next fundamental step was the discussion on $p$-adic dimensions
in physics started by Yu. Manin [59]. 

The important event in the $p$-adic world took place in 1987
when I. Volovich published  paper [60] on applications of $p$-adic numbers in
{\it{string theory.}} The string theory was new and rather intriguing attempt to reconsider 
foundations of physics by using space extended objects, strings, instead of the point weise objects, 
elementary particles. The scenarios of string spectacle is performed on fantastically 
small distances, so called {\it{Planck distances,}} $l_{P}\approx10^{-34} cm.$ Physicists have (at least) 
the feeling that space-time on Planck distances has some distinguishing features that could not be described by 
the standard mathematical model based on the field of real numbers ${\bf R}.$
In particular, there are ideas (that also are  strongly motivated by cosmology \footnote{We remark that one of 
the aims of string theory was to provide a new approach to general relativity. Therefore string investigations
are closely connected to investigations on fields of gravity and cosmology.}) that on Planck distances we could 
not more assume that there presents a kind of an order structure on the real line ${\bf R}.$
We remark that there is {\it{no order structure}} on ${\bf Q}_p$ (this is a disordered field). 

Another argument to consider a 
$p$-adic model of space-time on Planck distances is that ${\bf Q}_p$ is a {\it{non-Archimedean}} field. We do not plan 
to discuss here Archimedean axiom on the mathematical level of rigorousness. 

From the physical point of view
this axiom can be interpreted in the following way.
If we have some unit of measurement $l$, then we can measure each interval 
$L$ by using $l.$  By addition of this unit:
$l, l+l, l+l+l, \ldots, l+\ldots + l,$ we obtain 
larger and larger intervals that, finally, will cover $L.$
The precision of such a measurement is equal to $l.$
The process of such a type we can be realized 
in the field of real numbers ${\bf R}$. Therefore all physical 
models based on real numbers are Archimedean models. However, Archimedean axiom does not hold true in ${\bf Q}_p.$ 
Here successive addition does not increase the quantity. And there were  (long before $p$-adic physics)
intuitive cosmological ideas that space-time on 
Planck distances has non-Archimedean structure. 

In 80th and 90th there was demonstrated large interest
to various $p$-adic physical models, see, for example, papers on $p$-adic string theory of Aref'eva, Brekke, Dragovich, 
Framton, Freud, Parisi, Vladimirov, Volovich, Witten and many others, [60]-[64]
\footnote{We remark that these investigations in $p$-adic string theory were strongly based on the results
of mathematical investigations of 
I. M. Gelfand, M. I. Graev and I. I.  Pjatetskii-Shapiro
[44] on  distributions on $p$-adic fields.}, $p$-adic quantum mechanics and
field theory [65]-[70], $p$-adic models for spin glasses [71], [72]. These $p$-adic physical investigations stimulated 
the large interest to dynamical systems in fields of $p$-adic numbers ${\bf Q}_p$ and their finite and infinite extensions
(and, in particular, in the field of complex $p$-adic numbers ${\bf C}_p$). 

Continuous dynamical systems, namely $p$-adic differential 
equations, were studied by purely mathematical reasons. We can mention investigations of B. Dwork,
P. Robba, G. Gerotto, F. J. Sullivan [73]-[76], G. Christol [77], 
A. Escassut [25] on $p$-adic ordinary differential equations. However, $p$-adic physics stimulated investigations on new 
classes of continuous dynamical systems; in particular, partial differential equations over ${\bf Q}_p$, see, for example,
[78]-[83] ($p$-adic Schr\"odinger, heat, Laplace equations, Cauchy problem, distributions). We do not consider 
continuous $p$-adic dynamical systems in this review, see, e.g. book [69]. 

This review is devoted to discrete $p$-adic dynamical systems, 
namely iteration
\begin{equation}
x_{n+1}=f(x_n)
\end{equation}
of functions $f:{\bf Q}_p\rightarrow {\bf Q}_p$ or $f:{\bf C}_p\rightarrow {\bf C}_p.$
 
The development of investigations on $p$-adic discrete dynamical systems is the best illustration of
how physical models can stimulate new mathematical investigations. Starting with the paper on $p$-adic
quantum mechanics and string theory [67] (stimulated by investigations of Vladimirov and Volovich)  
E. Thiren, D. Verstegen and J. Weyers performed the first investigation 
on discrete $p$-adic dynamical systems [84] (iterations of quadratic polynomials). After this paper
investigations on discrete $p$-adic dynamical systems were proceeded in various 
directions [84]-[111], e.g. : 1) conjugate maps [84]-[86] and  [87] -- general technique based on Lie logarithms,
[96], [97] - problem of small denominators in ${\bf C}_p$; 2) ergodicity [98], [99], [97], [110]; 
3) random dynamical systems [100], [108]\footnote{See L. Arnold [112] for the general theory of random dynamical systems.} ; 
4) behaviour of cycles [101], [102], [106], [107], [111];
5) dynamical systems in finite extensions of ${\bf Q}_p$ [103], [104], [111]; 6) holomorphic and meromorphic
dynamics  [90], [26],  [93]-[95].

Recently discrete $p$-adic dynamical systems were applied to such
interesting and intensively developed domains as cognitive sciences and psychology.
[91], [113]-[122]. These cognitive applications are based on coding of human ideas by using branches of hierarchic 
$p$-adic trees and describing the process of thinking by iterations of $p$-adic dynamical system.
$p$-adic dynamical cognitive models were applied to such problems as memory recalling,
depression, stress, hyperactivity, unconscious and conscious thought and even Freud's psychoanalysis.
 These investigations stimulated the development of $p$-adic neural networks [116].
\footnote{We remark 
that in cognitive models there naturally appear $m$-adic trees for nonprime $m.$ Therefore
we also have to develop analysis and theory of dynamical systems in such a general case.}
Recently $p$-adic numbers were applied to problems of image recognition and compression
of information, see [123], [124].

\section{$p$-adic numbers}

The field of real numbers ${\bf R}$ is constructed as the completion of the
field of rational
numbers ${\bf Q}$ with respect to the metric $\rho(x,y)$ $=$ $\vert x - y
\vert$, where $\vert
\cdot\vert$ is the usual valuation  given by the absolute value.
The fields of $p$-adic
numbers ${\bf Q}_p$  are constructed in a corresponding way, but  using
other  valuations.
For a prime number $p$, the $p$-adic valuation $\vert \cdot\vert_p $ is
defined in the following
way. First we define it for natural numbers. Every natural number $n$ can
be represented as the
product of prime  numbers, $n$ $=$ $2^{r_2}3^{r_3} \cdots p^{r_p} \cdots$,
and we define
$\vert n\vert_p$ $=$ $p^{-r_p}$, writing $\vert 0 \vert_p$ $=0$  and
$\vert -n\vert_p$ $=$
$\vert n \vert_p$.  We then extend the definition of the $p$-adic valuation
$\vert\cdot\vert_p$
to all rational numbers by setting $\vert n/m\vert_p$ $=$ $\vert
n\vert_p/\vert m\vert_p$
for $m$ $\not=$ $0$. The completion of ${\bf Q}$ with respect to the metric
$\rho_p (x,y)$ $=$
$\vert x- y\vert_p$ is the locally compact field of $p$-adic numbers ${\bf
Q}_p$.

The number fields ${\bf R}$ and ${\bf Q}_p$ are unique in a sense, since by
{\it Ostrovsky's
theorem} [2] $\vert\cdot\vert$ and  $\vert\cdot\vert_p$ are the
only possible  valuations
on ${\bf Q}$, but have quite distinctive properties. The field of real
numbers ${\bf R}$ with its
usual valuation satisfies $\vert n\vert$ $=$ $n$ $\to$ $\infty$ for
valuations of natural numbers $n$
and is said to be {\it Archimedean.\/} By a well know theorem of number
theory [2] the only
complete Archimedean fields are those of the real and the complex numbers.
In contrast, the fields
of $p$-adic numbers, which satisfy $\vert n\vert_p$ $\leq$ $1$ for all $n$
$\in$ ${\bf N}$, are
examples of {\it non-Archimedean\/} fields. Here the Archimedean axiom is violated.
We could not get larger quantity by successive addition. Let $l$ be any  element 
of ${\bf Q}_p.$ There does not exist such a natural number $n$ that 
$\vert n l \vert_p\leq 1.$

The field of real numbers ${\bf R}$ is not isomorphic to any ${\bf Q}_p.$
Fields ${\bf Q}_s$ and ${\bf Q}_t$ are not isomorphic for $s\not=t.$
Thus starting with the field of rational numbers ${\bf Q}$ we get an infinite series of 
locally compact non-isomorphic fields:

${\bf Q}_2, {\bf Q}_3,..., {\bf Q}_{1997}, {\bf Q}_{1999},...$

Unlike the absolute value distance $\vert \cdot \vert$, the $p$-adic
valuation satisfies the strong
triangle inequality
\begin{equation}\label{str}
|x+y|_p \leq  \max[|x|_p,|y|_p],  \quad x,y \in {\bf Q}_p.
\end{equation}
Consequently the $p$-adic metric satisfies the strong triangle inequality
\begin{equation}
\label{s1}
\rho_p(x,y)\leq \max[\rho_p(x,z),\rho_p(z,y)], \quad x,y,z \in {\bf Q}_p,
\end{equation}
which means that the metric $\rho_p$ is an {\it ultrametric\/}.

Write $U_r(a)$ $=$ $\{x\in {\bf Q}_p: |x -a|_p \leq r\}$ and $U_r^-(a)$ $=$
$\{x\in {\bf Q}_p:
|x -a|_p < r\}$ where $r$ $=$ $p^n$ and $n$ $=$ $0$, $\pm 1$, $\pm 2$,
$\ldots$.
These are the ``closed'' and ``open''  balls in ${\bf Q}_p$ while the sets
$S _r(a)$ $=$ $\{x \in {\bf Q}_p: |x -a|_p = r \}$ are the spheres in ${\bf Q}_p$
of such radii $r$.
These sets (balls and spheres) have a somewhat strange topological
structure from the
viewpoint of our usual Euclidean intuition: they are  both open and closed at
the same time, and as such are called {\it clopen} sets.  Another
interesting property of $p$-adic
balls is that  two balls have nonempty intersection if and only if one of
them is contained
in the other. Also, we note that any point of  a $p$-adic ball can be
chosen as its center,
so such a ball is thus not uniquely characterized by its center and radius.
Finally, any $p$-adic
ball $U_r(0)$ is an additive subgroup of ${\bf Q}_p$, while the ball
$U_1(0)$ is also
a ring, which is called the {\it ring of $p$-adic integers} and is denoted
by ${\bf Z}_p$.

Any $x$ $\in$ ${\bf Q}_p$ has a unique  canonical expansion (which
converges in the
$\vert \cdot\vert_p$--norm) of the form
\begin{equation}
\label{a0}
x= a_{-n}/p^n +\cdots\ a_{0}+\cdots+ a_k p^k+\cdots
\end{equation}
where the $a_j$ $\in$ $\{0,1,\ldots, p-1\}$ are the ``digits'' of the
$p$-adic expansion and $n$ depend on $x.$ 

This expansion is similar to the standard expansion of a real number $x$
in the $p$-adic scale (e.g. binary expansion, $p=2):$
\begin{equation}
\label{A}
x=\cdots+ a_{-n}/p^n +\cdots\ a_{0}+\cdots+ a_k p^k
\end{equation}
In the $p$-adic case the expansion is finite in the direction of negative
powers of $p$ and infinite  in the direction of positive
powers of $p.$ In the real case the expansion is infinite in the direction of negative
powers of $p$ and finite  in the direction of positive powers of $p.$ 
In the $p$-adic case the expansion is unique; in the real case  - not.
The elements $x$ $\in$ ${\bf Z}_p$ have the expansion
$
x=a_{0}+\cdots+ a_k p^k+\cdots
$
and can thus be identified with the sequences of digits
\begin{equation}
\label{dit}
x = (a_0, ..., a_k,...).
\end{equation}

We remark that, as ${\bf Q}_p$ is a locally compact additive group, there 
exists the {\it Haar measure} $d m$ on the $\sigma$-algebra of Borel 
subsets of ${\bf Q}_p.$

If, instead of a prime number $p$, we start with an arbitrary natural number
$m > 1$ we construct the system of so-called $m$-adic numbers ${\bf Q}_m$
by completing ${\bf Q}$ with respect to the $m$-adic metric $\rho_m(x,y)$ $=$
$|x-y|_m$ which is defined in a similar way to above.  However, this system
is in general
not a field as there may exist divisors of zero; ${\bf Q}_m$ is only a
ring. Elements
of ${\bf Z}_m$ $=$ $U_1(0)$  can be identified with sequences (\ref{dit})
with the digits
$a_k$ $\in$ $\{0,1,\ldots,m-1\}$.  We can also use more  complicated
number systems corresponding to non-homogeneous scales $M$ $=$
$(m_1,m_2,...,m_k,...)$,
where the $m_j > 1$ are natural numbers, to obtain the number system
${\bf Q}_M$.  The elements $x$ $\in$ ${\bf Z}_M$ $=$ $U_1(0)$ can be
represented as sequences
of the form (\ref{dit}) with digits $a_j$ $\in$ $\{0,1,\dots, m_j-1\}$. The
situation
here becomes quite complicated  mathematically.  In general,  the number system
${\bf Q}_M$ is not a ring, but ${\bf Z}_M$ is always a ring.

We shall use a $p$-adic analogue of complex numbers. As we know, 
the field of complex numbers ${\bf C}$ is the quadratic extension of ${\bf R}$ with respect to 
the root of the equation $x^2+1=0: \; {\bf C}={\bf R}(i),
\; i=\sqrt{-1}, \; z=x+iy, \; x,y \in {\bf R}.$
In this case we have a very simple algebraic structure, because this quadratic extension 
is at the same time the algebraic closure of the field of real numbers
(every polynomial equation has a solution in ${\bf C}$).
In the $p$-adic case the structure of algebraic extensions is  more complicated.
A quadratic extension  is not unique.
If $p=2$ then there are seven quadratic extensions  and if $p\not =2,$
then there are three quadratic extensions. Thus if we consider the fixed quadratic
extension ${\bf Q}_p(\sqrt{\tau})$ of ${\bf Q}_p$ then there exist 
$p$-adic numbers for which it is impossible to find a square root in
${\bf Q}_p(\sqrt{\tau}).$ All quadratic extensions are not
algebraically closed. Extensions of any finite order 
(i.e, corresponding to roots of polynomials of any order) are not
algebraically closed. The algebraic closure ${\bf Q}_p^a$ of ${\bf Q}_p$ is constructed as an
infinite chain of extensions of finite orders. Therefore this is an
infinite-dimensional vector space over ${\bf Q}_p.$ This algebraic
closure is not a complete field. Thus we must consider the completion of this
field. It is the final step of this long procedure, because this
completion is an algebraically closed field (so we are lucky!),  Krasner's theorem, see e.g. [25].
Let us denote this field by ${\bf C}_p$. This field is called the field of complex
$p$-adic numbers.

\section{Roots of Unity}
The roots of unity in ${\bf C}_p$ will play an important role in our
considerations. To find fixed points and cycles of monomial 
functions $f(x)=x^n,$  we have to find the roots of unity.

As usual in arithmetics, $(n,k)$ denotes the greatest common divisor of
two natural numbers. Denote the group of $m$th roots of unity, $m=1,2,...,$ by 
$\Gamma^{(m)}.$ 
Set 
$$
\Gamma=\cup_{m=1}^\infty \Gamma^{(m)},\; \;
\Gamma_{m}=\cup_{j=1}^\infty \Gamma^{(m^j)}, \; \;
\Gamma_u= \cup_{(m,p)=1} \Gamma_m, 
$$
By elementary group theory we have $\Gamma =\Gamma_u \cdot \Gamma_p,
\; \Gamma_u\cap \Gamma_p=\{ 1\}.$ 

Denote the $k$th roots of unity by 
$\theta_{j,k}, \; j=1,...,k, \; \theta_{1,k}=1.$

We remark that $\Gamma_u \subset S_1(1)$ and $\Gamma_p \subset U_1^-(1)$.

The following estimate plays the important role in $p$-adic analysis
and theory of dynamical systems. We also present the proof to demonstrate
the rules of working in the framework of $p$-adic analysis. We denote binomial
coefficients by symbols 

$C_n^k= \frac{n!}{k!(n-k)!}, k\leq n.$

{\bf Lemma 3. 1.} {\it $|C_{p^k}^j|_p \leq 1/p$ for all $j=1,...,p^k-1.$}

{\bf Proof.} Let $j=ip +q, \; q=0,1,...,p-1.$ First consider the case $q=0:$
$$
\begin{array}{ll}
&
|C_{p^k}^j|_p=|\frac{p^k (p^k -p)\cdots (p^k-ip+p)}{p\cdots ip}|_p
\\ & \\
&
=\left| {p^k \over ip} {(p^k -p) \over p} \cdots {(p^k-ip+p)\over ip -p}\right|_p=
\left|{p^{k-1}\over i}\right|_p\leq {1\over p},
\\
\end{array}
$$
as $i<p^{k-1}.$ Now let $q \not= 0:$
$$
|C_{p^k}^j|_p =\left| p^k { (p^k -p)\over p} \cdots { (p^k-ip)\over
ip}\right|_p= |p^k|_p\leq {1\over p}\; .
$$\hfill\rule{2mm}{2mm} 

To find fixed points and cycles of functions $f(x)=x^n$ in ${\bf Q}_p,$ 
we have to know whether the roots of unity belong to ${\bf Q}_p.$ 
We present the corresponding result.
Denote by $\xi_l, \; l=1,2,...,$ a primitive $l$th root of 1 in ${\bf C}_p.$ We are
interested in whether $\xi_l \in {\bf Q}_p.$

{\bf Proposition 3. 1.} (Primitive roots) {\it If $p\not=2$ then 
$\xi_l \in {\bf Q}_p$ if and only if $l\;| \; (p-1).$ The field ${\bf Q}_2$
contains only $\xi_1=1$ and $\xi_2= -1.$}

To prove  this proposition we have to prove the same
result for the field $F_p=\{0,1,...,p-1\}$ of $\rm{mod}\; p$ residue classes 
and apply Hensel's lemma [2] ($p$-adic variant of Newton method, see appendix 1). This is 
one of the most powerful methods to get results for ${\bf Q}_p:$ first get such a result
for $F_p$ and try to find conditions to apply Hensel's lemma.

{\bf Corollary 3.1.} {\it The equation $x^k=1$ has $g=(k, p-1 )$ 
different roots in ${\bf Q}_p.$}

\section{Dynamical Systems in Non-Archimedean Fields}
To study dynamical systems in fields of $p$-adic numbers
${\bf Q}_p$ and complex $p$-adic numbers ${\bf C}_p$ as well as finite extensions
of ${\bf Q}_p,$
it is convenient to consider the general case of an arbitrary
non-Archimedean field $K.$

Let $K$ be a field (so all algebraic operations are well defined).
Recall that a {\it non-Archimedean valuation} 
is a mapping $|\cdot|_K:\; K\rightarrow {\bf R}_+$ satisfying 
the following conditions:
$
 |x|_K=0 \Longleftrightarrow x=0 \; \mbox{and}\; |1|_K=1;\;
\vert xy|_K = |x|_K|y|_K; \;
 |x+y|_K \leq \max(|x|_K, |y|_K). 
$
The latter inequality is the well known  strong triangle axiom.
We  often use in non-Archimedean investigations the following property of a
non-Archimedean valuation:

$
 |x+y|_F = \max(|x|_F, |y|_F) , \; \mbox{if} \; |x|_F \not= |y|_F. 
$

The field $K$ with the valuation $|\cdot|_K$ is called a 
non-Archimedean field. The fields of $p$-adic numbers
${\bf Q}_p$ and complex $p$-adic numbers ${\bf C}_p$ as well as finite extensions
of ${\bf Q}_p$ are non-Archimedean fields.

Thus all triangles in a non-Archimedean fields (in particular, in fields of $p$-adic numbers)
are {\it isosceles.}

Everywhere below  $K$ denotes   a   complete
non-Ar\-chi\-me\-de\-an field with a nontrivial valuation $| \cdot |_K$; $U_r(a), 
U_r^-(a)$ and $S_r(a)$
are respectively balls and spheres in $K.$ We always
consider $r \in |K|=\{ s=|x|_K: x\in K\}$ for radii  of balls  $U_r(a)$ and spheres
$S_r(a).$  In particular, in the $p$-adic case 
$r=p^{l}, l=0,\pm1, \pm 2,...$ and in the case of ${\bf C}_p$ - $r= p^q, q \in {\bf Q}.$

A function $f$ $:$ $U_r(a)$ $\to$ $K$  is said to be {\it
analytic\/} if it can be  expanded
into a power series $f(x)$ $=$ $\sum_{n=0}^\infty f_n (x - a)^n$ with $f_n$
$\in$ $K$ which converges  uniformly on the ball $U_r(a)$.

Let us study the dynamical system:
\begin{equation}
\label{2.1}
U \to U, \; \; \; x \to f(x), 
\end{equation}
where  $U=U_R(a)$ or $K$ and $f:U \to U$
is an analytic function. First we shall prove a general theorem about a
behaviour of iterations $ x_n = f^n(x_0), \;  x_0 \in U.$ As usual
$f^n(x)=f \circ ...\circ f(x).$ Then we shall
use this result to study a behaviour of the concrete dynamical systems
$f(x)=x^n, \; n=2,3,...,$ in the fields of complex $p$-adic numbers ${\bf C}_p.$

We shall use the standard terminology of the theory of dynamical
systems. 
If $f(x_0)=x_0$ then $x_0$ is a {\it fixed point.}
If $x_n=x_0$  for some $n=1,2,...$ we say that $x_0$ is a {\it periodic
point}. If $n$ is the smallest natural number with this property then $n$
is said to be the {\it period} of $x_0.$ We denote the corresponding {\it cycle} 
by $\gamma=(x_0, x_1,...,x_{n-1}).$ In particular, the fixed point $x_0$ is the 
periodic point of period 1. Obviously $x_0$ is a fixed point of the iterated
map $f^n$ if $x_0$ is a periodic point of period $n.$ 

A fixed point $x_0$ is 
called an
{\it attractor} if there exists a neighborhood $V(x_0)$ of $x_0$ such that
all points $y \in V(x_0)$ are attracted by $x_0,$ i.e.,
$\lim_{n \to \infty} y_n= x_0.$ If $x_0$ is an attractor, we consider
its {\it basin of
attraction} $A(x_0)= \{ y \in K : y_n \to x_0, n\to \infty\}.$ A fixed point
$x_0$ is called {\it repeller} if there exists a neighborhood $V(x_0)$ of $x_0$ 
such that $|f(x) -x_0|_K > |x - x_0|_K$ for $x \in V(x_0), x \not=x_0.$
A cycle $\gamma=(x_0, x_1,...,x_{n-1})$ is said to be an attractor (repeller)
if $x_0$ is attractor (repeller) of the map $f^n.$

We have to be more careful
in defining a non-Archimedean analogue of a {\it Siegel disk.}
In author's book [91] non-Archimedean Siegel disk
was defined in the following way.
Let $a \in U$ be a fixed point of a function 
$f(x).$ The ball $ U_r^-(a)$ (contained in $U$)
is said to be a Siegel disk 
if each sphere $S_\rho(a), \; \rho < r,$ is an invariant sphere of
$f(x),$ i.e.,
if one takes an initial point on one of the spheres $S_\rho(a), \; \rho<r,$
all iterated points will also be on it. The union of all Siegel disks 
with center in $a$ is said to be a maximal Siegel disk. Denote the
maximal Siegel disk by $SI(a).$

{\bf Remark 4.1.} {\small In complex geometry the center of a disk is uniquely 
determined by the disk. Hence it does not happen that different fixed points have the 
same Siegel disk. But in non-Archimedean geometry centers of a disk are 
nothing but the points which belong to the disk. And in principle different fixed
points may have the same Siegel disk (see the next section).}

In the same way we define a Siegel disk with center at a periodic point 
$a \in U$ with the corresponding cycle $\gamma=\{ a, f(a),...,
f^{n-1}(a)\}$ of the period $n.$ Here the spheres $S_\rho(a),
\; \rho<r,$ are invariant spheres of the map $f^n(x).$ 

As usual in the theory of dynamical systems, we can find attractors,
repellers, and Siegel disks using  properties of the derivative of
$f(x).$ Let $a$ be a periodic point with period $n$
of $C^1$-function $g: U \to U.$ Set 
$\lambda=\frac{d g^n (a)}{d x} .$ 
This point is called: 1) {\it attractive}  if 
$0 \leq \vert \lambda \vert_K <1; \; \; \;$ 2) {\it indifferent}
if $\vert \lambda \vert_K =1;$ 3) {\it repelling}  if
$\vert \lambda \vert_K > 1.$

{\bf Lemma 4.1.} [91] {\it Let $f:U\to U$ be an analytic function and let $a\in U$ and $f^\prime(a)\not=0.$ 
Then there exist $r>0$ such that
\begin{equation}
\label{2.2}
s=\max_{2\leq n<\infty}\left\vert \frac{1}{n!}\frac{d^n f}{d x^n}(a)\right\vert_K r^{n-
1} < \vert f^\prime(a)\vert_K. 
\end{equation}
If $r>0$ satisfies this inequality and $U_r(a) \subset U$ then 
\begin{equation}
\label{2.3}
\vert f(x) - f(y)\vert_K = \vert f^\prime(a)\vert_K \vert x - y \vert_K 
\end{equation}
for all $x,y\in U_r(a).$}

By using the previous Lemma we prove:

{\bf Theorem 4.1.} {\it Let $a$ be a fixed point of the analytic function
$f:U \to U.$ Then:

{\rm{1.}} If $a$ is an attracting point of $f$ then it is an attractor of the
dynamical system (\ref{2.1}). If $r>0$ satisfies the inequality:
\begin{equation}
\label{2.5}
q=\max_{1\leq n<\infty }\left\vert \frac{1}{n!}\frac{d^n f}{d x^n}(a)\right\vert_K r^{n-
1} < 1, 
\end{equation}
and $U_r(a) \subset U$ then $U_r(a) \subset A(a).$

{{\rm{2.}}} If $a$ is an indifferent point of $f$
then it is the center of a Siegel disk.  If $r>0$ satisfies the inequality
(\ref{2.2}) and $U_r(a) \subset U$  then $U_r(a)\subset SI(a).$

{\rm{3.}} If $a$ is a repelling point of $f$ then $a$ is a repeller of the dynamical 
system (\ref{2.1}).}

We note that (in the case of an attracting point) the condition (\ref{2.5}) 
is less restrictive than the condition (\ref{2.2}).

To study dynamical systems for nonanalytic functions we can use the 
following theorem of non-Archimedean analysis [23]:

{\bf Theorem 4.2.} (Local injectivity of $C^1$-functions) {\it Let
$f: U_r(a)\to K$ be $C^1$ at the point $a.$ If $ f^\prime(a)\not=0$
there is a ball $U_s(a) , \; s\leq r,$ such that (\ref{2.3})
holds for all $ x,y \in U_s(a).$}

However, Theorem 4.1 is more useful for our considerations,
because Theorem 4.2 is a so-called `existence theorem'.
This theorem does not
say anything about the value of $s.$ Thus we cannot estimate
a volume of $A(a)$ or $SI(a).$ Theorem 4.1 gives us such a
possibility. We need only to test one of the conditions (\ref{2.5}) or
(\ref{2.2}). 

Moreover, the case $f^\prime(a)=0$ is `a pathological case'
for nonanalytic functions of a non-Archimedean argument. For example,

{\it there exist functions $g$ which are not locally constant but
$g^\prime = 0$ in every point.}

In our analytic framework we have no such problems.

A {\it Julia set} $J_f$ for the dynamical system (\ref{2.1}) is defined as 
the closure of the set of all repelling periodic points of 
$f.$ The set $F_f=U\setminus J_f$ is called a {\it Fatou set.}
These sets play an important role in the theory of real dynamical
systems. In the non-Archimedean case the structures of these sets were investigated
in [91], [93], [95].

We shall also use an analogue of Theorem 4.1 for periodic points. There 
we must apply our theorem  to the iterated function $f^n(x).$

\section{Dynamical Systems in the Field of Complex $p$-adic Numbers} 

As an application of Theorem 4.1 we study the simplest discrete dynamical systems,
namely monomial systems:

$f(x)=\psi_n(x)=x^n, \; n=2,3,...,$ 

in fields of complex $p$-adic numbers ${\bf C}_p.$ We shall see that behaviour of
$\psi_n$ crucially  depends on the prime number $p,$ the base of the corresponding
field. Depending on $p$ attractors and Ziegel disks appear and disappear
transforming one into another. Especially complex dependence on $p$ will be studied
in section 6 devoted to dynamical systems in ${\bf Q}_p, p=2,3,..., 1997, 1999,...$

It is evident that the points $a_0$ and $a_\infty$ are
attractors with basins of attraction $A(0)=U_1^-(0)$ and 
$A(\infty)={\bf C}_p \setminus U_1(0),$ respectively. Thus the main
scenario is developed on the sphere $S_1(0).$ Fixed points of $\psi_n(x)$ 
belonging to this sphere are the roots $\theta_{j,n-1}, \; j=1,...,n-1,$
of unity of degree $(n-1).$ There are two essentially different cases: 
1) $n$ is not divisible by $p;$ 2) $n$ is divisible by $p.$ 
The proof of the following theorem is based on the results of Theorem 4.1.

{\bf Theorem 5.1.} {\it The dynamical system $\psi_n(x)$ has $(n-1)$ fixed points 
$a_j=\theta_{j,n-1}, \; j=1,...,n-1,$ on the sphere $S_1(0).$

1. Let $(n,p)=1.$ There all these points are centers of Siegel disks and $SI(a_j)=
U_1^-(a_j).$ If $n-1= p^l, \; l=1,2,...,$ then $SI(a_j)=SI(1)=U_1^-(1)$
for all $j=1,...,n-1.$ If $(n-1, p)=1,$ then $a_j\in S_1(1), \;
j=2,...,n-1,$ and $SI(a_j)\cap SI(a_i)=\emptyset, \; i\not=j.$ For any $k=2,3,...$ all $k$-cycles
are also centers of Siegel disks of unit radius.

2. If $(n,p)\not= 1,$ then these points are attractors and  
$U_1^-(a_j) \subset A(a_j).$ For any $k=2,3,...$ all $k$-cycles are also 
attractors and open unit balls are contained in basins of attraction.}

{\bf Corollary 5.1.} {\it Let $(n,p)=1.$ Let $n^k-1=p^l, \; l=1,2,...$
Any $k$-cycle $\gamma=(a_1,...,a_k)$
for such a $k$ is located in the ball $U_1^-(1);$ it has the behaviour of a
Siegel disk with $SI(\gamma)=\cup_{j=1}^k U_1^-(a_j)= U_1^-(1).$
During the process of the motion the distances $c_j=\rho_p(x_0,a_j),
\; j=1,...,k,$ where $x_0 \in U_1^-(1)$ is an arbitrary initial point,
are changed according to the cyclic law: $(c_1,c_2,...,c_{n-1},c_n)
\to (c_{n},c_1,...,c_{n-2},c_{n-1}) \to ...\; .$}

Thus in the case $(n,p)=1$ the motion of a point in the ball $U_1^-(1)$
is very complicated. It moves cyclically (with different periods) around an infinite number of centers.
Unfortunately, we could not paint a picture of such a motion in our Euclidean space -
it is too restrictive for such images.

Theorem 5.1 does not completely describe the case $(n,p)=1, (n-1,p)\not=1.$
Let us consider the general case: $(n,p)=1, n-1= m p^l$ with $(m,p)=1$ and 
$l\geq 0.$ Set $a_i=\xi_m^i, i=0,1,..., m-1,\; b_j=\xi_{p^l}^j, j=0,1,...,
p^l-1,$ and $c_{ij}=a_i b_j.$ Then all these points $c_{ij}, i=0,1,..., m-1,
j=0,1,...,p^l-1,$ are centers of Siegel disks and $SI(c_{ij})=
U_1^-(c_{ij}).$ For each $i$ we have $SI(c_{i0})=SI(c_{i1})=...=
SI(c_{i(p^l-1)}).$ If $i \not=0$ then all these disks $SI(c_{ij})$ are in $S_1(0)
\cap S_1(1).$ Further, $SI(c_{ij}) \cap SI(c_{kl})=\emptyset$ if $i\not=k.$
We can formulate the same result for $k$-cycles.

Now we find the basins of attraction $A(a_j), \; j=1,..., n-1, \; (n,p)\not=1,$ exactly. 
We begin from the attractor $a_1=1.$ 

\medskip

Let $n= m p^k, \; (m,p)=1,$ and $k\geq 1.$

{\bf Lemma 5.1.} {\it The basin of attraction $A(1)=
\cup_{\xi}U_1^-(\xi)$ where $\xi\in \Gamma_m ;$
these balls have empty intersections for different points $\xi.$}

{\bf Corollary 5.2.} {\it Let $n=p^l, l\geq 1.$ Then $S_1(1)$ is an 
invariant sphere of the dynamical system $\psi_n(x).$}

{\bf Examples.} 1. Let $n=p^l, \; l\geq 1.$ Then $A(1)=U_1^-(1).$

2. Let $p\not=2$ and $n=2 p^l, \; l\geq 1.$ Then $A(1)= \cup U_1^-(\xi)$ 
where $\xi\in \Gamma_2.$

{\bf Theorem 5.2.} {\it The basin of attraction $A(a_k)=
\cup_\xi U_1^-(\xi a_k)$ where $\xi \in \Gamma_m.$ 
These balls have empty intersections for different points $\xi.$}

The dynamical system $\psi_n(x)$ has no repelling points in ${\bf C}_p$ for any $p.$ Thus
the Julia set $J_{\psi_n}=\emptyset$ and the Fatou set ${\cal F}_{\psi_n}={\bf C}_p,$  cf. [93], [95].

\section{Dynamical Systems in the Fields of $p$-adic $\; \; \; \; $
 Numbers}

Here we study the behaviour of the dynamical system $\psi_n(x)= x^n, \;
n=2,3,...,$
in ${\bf Q}_p.$ In fact, this behaviour can be obtained on the basis 
of the corresponding behaviour in ${\bf C}_p.$ We need only to apply the results of  section 3
about the roots of unity in ${\bf Q}_p$

{\bf Proposition 6.1.} {\it The dynamical system $\psi_n(x)$ has $m=(n-1,p-1)$ fixed points 
$a_j= \theta_{j,m}, j=1,...,m,$ on the sphere $S_1(0)$ of ${\bf Q}_p.$ The character of 
these points is described by Theorem 5.1.
Fixed points $a_j\not=1$ belong to the sphere $S_1(1).$ }

We remark that a number of attractors or Siegel disks for the dynamical system
$\psi_n(x)$ on the sphere $S_1(0)$ is $\leq (p-1).$

To study $k$-cycles in ${\bf Q}_p$ we use the following numbers: 

$m_k=(l_k, p-1), \; k=1,2,...,$ 
with $l_k=n^k -1.$ 

{\bf Proposition 6.2.} {\it The dynamical system $\psi_n(x)$ has $k$-cycles ($k\geq 2$) in 
${\bf Q}_p$ if and only if $m_k$ does not divide any $m_j, \; j=1,...,k-1.$
All these cycles are located on $S_1(1).$}

In particular, if $(n,p)=1$ (i.e., all fixed points and $k$-cycles are centers of Siegel disks) 
there are no such complicated motions around a group of centers as
in ${\bf C}_p.$

{\bf Corollary 6.1.} {\it The dynamical system $\psi_n(x)$ has only a finite number of cycles
in ${\bf Q}_p$ for any prime number $p.$}

{\bf Example.} Let $n=p^l, \; l\geq 1.$ Then $m_1= p-1$ and there are $p-1$ attractors $a_j=\theta_{j,p-1}, \; j=1,...,p-1,$ with the basins of attraction
$A(a_j)=U_{1/p}(a_j)$ and there is no $k$-cycle for $k\geq2.$
 As we can choose $a_j=j \; \rm{mod}\; p,$ then 
$U_{1/p}(a_j)=U_{1/p}(j).$ In particular, if $p=2$ then
all points of the sphere $S_1(0)$ are attracted by $a_1.$

To study the general case $n=q p^l, l\geq 1, (q,p)=1,$ we use the following elementary fact.

{\bf Lemma 6.1.} {\it Let $n=q p^l, \; l\geq 1, \; (q,p)=1.$
Then $m_k=(l_k,p-1)=(q^k-1,p-1), \; k=1,2,... \;.$}

{\bf Examples.} 1). Let $n=2p, \; p\not=2.$ There is only one attractor
$a_1=1$ on $S_1(0)$ for all $p.$ 
To find $k$-cycles, $k\geq 2,$ we have to consider the  numbers
$m_k, \; k=2,... \;.$ However, by Lemma 6.1 $m_k=(2^k-1, p-1).$ Thus the number
of $k$-cycles for the dynamical system $\psi_{2p}(x)$ coincides with the corresponding 
number for the dynamical system $\psi_2(x).$ An extended analysis
of the dynamical system $\psi_2(x)$ will be presented after Proposition 6.3. Of
course, it should be noted that the behaviours of $k$-cycles for $\psi_{2p}(x)$ and 
$\psi_2(x), \; p\not=2,$ are very different. In the first case these are attractors;
in the second case these are centers of Siegel disks.

2). Let $n=3p, \; p\not=2.$ Then there are two attractors $a_1=1$ and $a_2=-1$ on $S_1(0)$ for all $p.$ 

3). Let $n=4p.$ 
Here we have a more complicated picture: 1 attractor for
$p=2,3,5,11,17,23,...;$
3 attractors for $p=7,13,19,29,31,... \;.$ 

4). Let $n=5p.$ Here we have: 1 attractor for $p=2;$ 2 attractors for $p=3,7,11,23,31,...;$   4 attractors for
$p=5,13,17,...\; .$

We now study basins of attraction 
(in the case $n=q p^l, \; l \geq 1, \; (q,p)=1).$ As a consequence of our investigations for the dynamical system in ${\bf C}_p$ we find that $A(1)=
\cup_\xi U_{1/p}(\xi)$ where $\xi \in \Gamma_q \cap {\bf Q}_p.$ We have
$\Gamma_q \cap {\bf Q}_p\not=\{1\}$ iff $(q,p-1)\not=1.$

{\bf Examples.} 1). Let $p=5$ and $n=10,$ i.e., $q=2.$ As $(q^2,p-1)=4,$ 
then $\Gamma_2 \cap {\bf Q}_5=\Gamma^{(4)}$ and $A(1)=\cup_{j=1}^4 
U_{1/5}(\theta_{j,4}).$ Thus $A(1)=S_1(0).$ All points of the sphere
$S_1(0)$ are attracted by $a_1=1.$ 

2). Let $p=7$ and $n=21,$ i.e., $q=3.$ There $m_1=(q-1, p-1)=2.$ Hence there
are two attractors; these are $a_1=1$ and $a_2=-1.$ As all $m_j=(q^k-1,p-1)=2,
\; j=1,2,...,$ then there are no $k$-cycles for $k\geq 2.$ 

3). Let $p=7$ and $n=14,$ i.e., $(q,p-1)=2.$
Thus $\Gamma_2 \cap {\bf Q}_7=\Gamma^{(2)}$ and 
 as $m_2=3,$ there exist 2-cycles. It is easy to see that the 2-cycle is unique and
$\gamma=(b_1,b_2)$ with $b_1=2,b_2=4 \; \rm{mod} \; 7.$ This cycle
generates a
cycle of balls on the sphere $S_1(1): \; \gamma^{(f)}=(U_{1/7}(2), U_{1/7}(4))$ (`fuzzy cycle').
Other two balls on $S_1(1):\; U_{1/7}(3), U_{1/7}(5)$ are attracted
by $\gamma^{(f)}$ (by the balls $U_{1/7}(2)$ and $U_{1/7}(4),$ respectively).

The last example shows that sometimes it can be interesting to study not
only cycles of points but also cycles of balls. We propose the following
general definition, see [91]:

Let $x\to g(x) , \; x \in {\bf Q}_p ,$ be a dynamical system. If there exist
balls $U_{r}(a_j), \; j=1,...,n,$ such that  iterations of the
dynamical system generate the cycle of balls $\gamma^{(f)}=
(U_{r}(a_1),...,U_{r}(a_n )),\; (r=p^{l}, l=0,\pm 1,...)$  then it is
called a {\it fuzzy cycle} of length $n$ and radius $r.$ Of course,
we assume that the balls in the fuzzy cycle  do not coincide.

{\bf Proposition 6.3.} {\it There is a one to one correspondence between
cycles and fuzzy cycles of radius $r=1/p$ of the dynamical system $\psi_n(x)$ 
in ${\bf Q}_p.$}

The situation with fuzzy cycles of radius $r<1/p$ is more complicated,  see the following example 
and appendix 2.

{\bf Examples.} Let $p=3,n=2.$ There exist 2-cycles of radius $r=1/9$
which do not correspond to any ordinary cycle. For example, $\gamma^{(f)}=(4,7).$ Further,
there exist fuzzy 6-cycles with $r=1/27;$ fuzzy 18-cycles with
$r=1/81,$... .

We now consider prime-number dependence of cyclic behaviour for the dynamical
system: $\psi_2(x)= x^2.$ This is the simplest among monomial dynamical systems.
However, even here we observe very complicated dependence on $p.$

{\bf Examples.} Let $n=2$ in all the following examples.

1). Let $p=2.$ There is only one fixed point $a_1=1$ on $S_1(0).$
It is an attractor and $A(1)=U_{1/2}(1)=S_1(0).$

2). As $l_k$ are odd numbers, then $m_k$ must also  be an odd number. Therefore there are no
any $k$-cycle ($k>1$) for $p=3,5,17$ and for any prime number 
which has the form $p=2^k+1$. 

3). Let $p=7.$ Here $m_k$ can be equal to 1 or 3. As $m_2=3$ there are only 2-cycles. It is easy 
to show that the 2-cycle is unique. 

4). Let $p=11.$ Here $m_k=1$ or 5. As $m_2=m_3=1$ and $m_4=5$ there exist only 4-cycles. There 
is only one 4-cycle: $\gamma(\xi_5).$

5). Let $p=13.$ Here $m_k=1$ or 3. As $m_2=3$ there exists only the (unique) 2-cycle.

6). Let $p=19.$ Here $m_k=1$ or 3, or 9. As $m_2=3$ there is the (unique) 2-cycle. However,
although $m_4=3$ there are no 4-cycles because $m_4$ divides $m_3.$ As $m_6=9$ does not 
divide $m_2,...,m_5$ there exist 6-cycles and there are no $k$-cycles with $k>6.$
There is only one 6-cycle: $\gamma(\xi_9).$

7). Let $p=23.$ Here $m_k=1$ or 11. The direct computations show that there are no $k$-cycles 
for the first $k=2,...,8.$ Further computations are complicated. We think that an answer
to the following question must be  known in number theory: Does there exist $k$ such that 11 divides $l_k?$

8). Let $p=29.$ Here $m_k=1$ or 7. As $m_3=7$ and $m_2=1$ there exist only 3-cycles. It is easy
to show that there are two 3-cycles: $\gamma(\xi_7)$ and $\gamma(\xi_7^3).$

9). Let $p=31.$ Here $m_k=1, 3,5,15.$ As $m_2=3$ there exists an (unique) 2-cycle. 
As $m_4=15$ and $m_3=1$ there exist 4-cycles: $\gamma(\xi_{15}), \gamma(\xi_{15}^3),$ $\gamma(\xi_{15}^7).$ 
There are no $k$-cycles with $k\not= 2,4.$

10). Let $p=37.$ Here $m_k=1,3,9.$ As $m_2=3$ there exists an (unique) 2-cycle.
As $m_6=9$ and $m_2=m_4=3, \; m_3=m_5=1$ there exist 6-cycles. It is easy to show that
there is an unique 6-cycle: $\gamma(\xi_9).$ There are no $k$-cycles for $k\not=2,6.$

11). Let $p=41.$ Here $m_k=1,5.$ As $m_4=5$ and all previous $m_j=1$ there exist
4-cycles. It is easy to show that this cycle is unique: $\gamma(\xi_5).$ There are no
$k$-cycles with $k\not=4.$

Thus, even for the simplest $p$-adic dynamical system, $x\to x^2,$ 
the structure of cycles depends in a very complex way on the parameter $p.$
In general case this dependence was studied in [101], [102], [106]. It was demonstrated 
that a number ${\cal N}_m$ of cycles of the fixed length, $m,$ depends randomly on $p.$
So ${\cal N}_m(p)$ is a random variable defined on the set of all prime numbers.
We found mean value and covariance of this random variable [101], [102], [106]. There we used
essentially classical results on the distribution of prime numbers, see e.g.  [125], [126].
We hope that the connection between the theory of $p$-adic dynamical
systems and the classical theory
on distributions of prime numbers established in  [101], [102], [106] will have
further applications.

\section{$p$-adic ergodicity}

In this section we study in details ergodic behavior of $p$-adic monomial
dynamical systems. As we have already seen in previous section, behaviour 
of $p$-adic dynamical systems depends crucially on the prime parameter
$p.$ The main aim of investigations performed in paper [98] was to find such a 
$p$-dependence for ergodicity.

Let $\psi_n$ be a (monomial) mapping on ${\bf Z}_p$ taking $x$ to $x^n$. 
Then all spheres
$S_{p^{-l}}(1)$ are 
$\psi _n$-invariant iff $n$ is a multiplicative unit, i.e., $(n,p)=1.$ 

In particular $\psi_n$ is an isometry on $S_{p^{-l}}(1)$ if and only if $(n,p)=1$.
Therefore we will henceforth
assume that $n$ is a
unit.  Also note that, as a consequence, $S_{p^{-l}}(1)$ is not a group under 
multiplication. Thus our investigations are {\it not about the dynamics on a 
compact (abelian) group. }

We remark that monomial mappings, $x\mapsto x^n$, are topologically transitive and
ergodic with respect to Haar measure on the unit circle in the
complex plane.
We obtained [98] an analogous result for monomial dynamical systems
over $p-$adic
numbers. The process is,
however, not straightforward. The result will depend on the natural number
$n$. Moreover,
in the $p-$adic case we never have ergodicity on the unit circle, but on
the circles
around the point $1$.

{\bf 7.1. Minimality.}
Let us consider the dynamical system $x\mapsto x^n$ on spheres 
$S_{p^{-l}}(1)$. 
The result depends crucially on the following well known result from group 
theory. We denote the multiplicative group of the ring $F_k$ of $\rm{mod}\; k$
residue classes by the symbol $F^\star_k;$  we also set $<n>= \{ n^N: N=0,1,2,...\}$
for a natural number $n.$
 
{\bf Lemma 7.1.}
{\it Let $p>2$ and $l$ be any natural number, then the natural number $n$ is a 
generator of
$F^\star_{p^l}$ if and only if $n$ is a generator of $F^\star_{p^2}$. $F^\star_{2^l}$ is 
noncyclic for
$l\geq 3$.}

\vspace{1.5ex}
\noindent
Recall that a dynamical system given by a continuous transformation $\psi$ on
a compact metric space $X$ is called \emph{topologically transitive} if there 
exists
a dense orbit $\{\psi^n(x): n\in {\bf N}\}$ in $X$, and (one-sided)
\emph{minimal}, if all orbits for $\psi$ in $X$ are dense. For the case of 
monomial systems $x\mapsto x^n$ on spheres $S_{p^{-l}}(1)$ topological 
transitivity
means the existence of an $x\in S_{p^{-l}}(1)$ s.t. each $y\in S_{p^{-l}}(1)$
is a limit point in the orbit of $x$, i.e. can be represented as
\begin{equation}\label{limit point}
y=\lim_{k\rightarrow\infty} x^{n^{N_k}},
\end{equation}
for some sequence $\{N_k\}$, while minimality means that such a property holds
for any $x\in S_{p^{-l}}(1)$. Our investigations are based on the following 
theorem.

{\bf Theorem 7.1.}
    {\it For $p\neq 2$ the set $\langle n\rangle$ is dense in
    $S_1(0)$ if and only if $n$ is a generator of $F^\star_{p^2}$. }
  
    {\noindent \bf Proof. }%
      We have to show that for every $\epsilon >0$ and 
      every $x\in S_1(0)$ there is a $y\in \langle n\rangle$ such 
      that $\left|x-y\right|_p<\epsilon$. Let $\epsilon>0$ and $x\in S_1(0)$ be
      arbitrary. Because of the discreteness of the $p-$adic metric we can assume that
      $\epsilon=p^{-k}$ for some natural number $k$. But (according to Lemma 7.1)
      if $n$ is a generator of $F^\star_{p^2}$,
      then $n$ is also a generator of $F^\star_{p^l}$ for every natural number $l$ (and $p\neq 2$)
      and especially
      for $l=k$. Consequently there is an $N$ such that $n^N=x \;\rm{mod}\; p^k$. From the
      definition of the $p-$adic metric we see that $\left|x-y\right|_p<p^{-k}$ if and only if
      $x$ equals to $y \; \rm{mod}\; p^k$. Hence we have that $\left|x-n^N\right|_p<p^{-k}$.

Let us consider $p\not=2$ and for $x\in U_{p^{-1}}(1)$ the $p$-adic 
exponential  function $t \mapsto x^t,$ see, for example [23].
This function is well defined and continuous as a map from ${\bf Z}_p$ to
${\bf Z}_p$.
In particular, for each $a \in {\bf Z}_p$, we have 
\begin{equation}
\label{l1}
x^a=\lim_{k\rightarrow a} x^k, \; \; k\in {\bf N}.
\end{equation}
We shall also use properties of the $p$-adic logarithmic function, see, for
example [23].
Let $ z \in U_{p^{-1}}(1).$  Then $\log z$ is well defined. 
For $z=1+\lambda$ with $\left|\lambda \right|_p \leq 1/p$, we have:
\begin{equation}
\label{l2}
 \log{z}=\sum_{k=1}^{\infty} \frac{(-1)^{k+1}\Delta^k}{k}=\lambda
(1+\lambda \Delta_\lambda),
      \quad \left|\Delta_\lambda \right|_p\leq 1.
\end{equation}      

\vspace{1.5ex}
\noindent  
By using (\ref{l2}) we obtain that  $\log : U_{p^{-1}}(1) \to U_{p^{-1}}(0)$ is
an isometry:
\begin{equation}
\label{l3}
\vert \log x_1 - \log x_2 \vert_p= \vert x_1 - x_2\vert_p,\; \;  x_1, x_2
\in U_{1/p}(1) \; .
\end{equation}

{\bf Lemma 7.2.} {\it Let $x \in U_{p^{-1}}(1), x \not = 1, a \in {\bf Z}_p$ and 
let $\{ m_k \}$ be a sequence of natural numbers. If $x^{m_k} \to x^a,
k \to \infty,$ then $m_k \to a$ as $k \to \infty,$ in ${\bf Z}_p$.}

This is a consequence of the isometric property of $\log$.

{\bf Theorem 7.2.}
    {\it Let $p\not=2$ and $l\geq 1$. Then the monomial dynamical system
    $x\mapsto x^n$ is minimal on the circle $S_{p^{-l}}(1)$ if and only if 
$n$ is a
    generator of $F^\star_{p^2}$.}
  
{\noindent \bf Proof. }%
Let $x \in S_{p^{-l}}(1).$
Consider the equation $x^a = y$. What are the possible
values of $a$ for $y \in S_{p^{-l}}(1)$?  
We prove that $a$ can take an arbitrary  value from the sphere
$S_1(0).$ We have that $a=\frac{\log x}{\log y}$. 
As $\log: U_{p^{-1}}(1) \to U_{p^{-1}}(0)$ is an isometry, we
have $\log(S_{p^{-l}}(1)) = S_{p^{-l}}(1).$ Thus 
$a= \frac{\log x}{\log y} \in S_1(0)$ and moreover, each
$a \in S_1(0)$ can be represented as $\frac{\log x}{\log y}$
for some $y \in S_{p^{-l}}(1)$.

Let $y$ be an arbitrary element of $S_{p^{-l}}(1)$ and 
let $x^a=y$ for some $a\in S_1(0).$ By Theorem 7.1 if $n$ is a generator of
$F^\star_{p^2},$ 
then each $a \in S_1(0)$ is a limit point of the sequence $\{ n^N
\}_{N=1}^\infty.$
Thus $a= \lim_{k\to\infty} n^{N_k}$ for some subsequence $\{N_k \}.$ By
using the continuity
of the exponential function we obtain (\ref{limit point}).

Suppose now that, for some $n,$ $x^{n^{N_k}} \to x^a.$ By Lemma 7.2 we
obtain that
$n^{N_k} \to a$ as $k \to \infty.$ If we have (\ref{limit point}) for all $y
\in S_{p^{-l}}(1),$
then each $a\in S_1(0)$ can be approximated by elements $n^N.$ In
particular,
all elements $\{1,2,...,p-1,p+1,.., p^2-1 \}$ can be approximated with
respect to $\rm{mod}\; p^2.$
Thus $n$ is a is a generator of $F^\star_{p^2},$

\vspace{1.5ex}
\noindent {\bf Example. }In the case that $p=3$ we have that $\psi_n$ is 
minimal if $n=2$,
$2$ is a generator of $F^\star_{9}=\{1,2,4,5,7,8\}$. But for $n=4$ it is not; 
$\langle 4\rangle \rm{mod}\; 3^2=\{1,4,7\}$. We can also see this by noting that
$S_{1/3}(1)=U_{1/3}(4)\cup U_{1/3}(7)$ and that $U_{1/3}(4)$ is invariant 
under $\psi_4$.

{\bf Corollary 7.1.}
{\it If $a$ is a fixed point of the monomial dynamical system $x\mapsto x^n$, then
this is minimal on $S_{p^{-l}}(a)$ if and only if $n$ is a generator of 
$F^\star_{p^2}$.}

\vspace{1.5ex}
\noindent {\bf Example. }
Let $p=17$ and $n=3$. In ${\bf Q}_{17}$ there is a primitive $3$rd root of 
unity, see for example [2]. 
Moreover, $3$ is also a generator of $F^\star_{17^2}$. Therefore there exist
$n$th roots of unity different from $1$ around which the dynamics is
minimal.

{\bf 7.2. Unique ergodicity.}
In the following we will show that the minimality of the monomial dynamical
system $\psi_n:$ $x\mapsto x^n$ on the sphere $S_{p^{-l}}(1)$ is equivalent 
to its \emph{unique ergodicity}. The latter property means that there exists 
a unique probability measure on $S_{p^{-l}}(1)$ and its Borel $\sigma-$algebra
which is invariant under $\psi_n$. We will see that this 
measure is in fact the normalized restriction of the Haar measure on
${\bf Z}_p$. Moreover, we will also see that the ergodicity of $\psi_n$ with
respect to Haar measure  is also equivalent to its unique
ergodicity. We should point out that -- though many results are
analogous to the case of the (irrational) rotation on the circle, our
situation is quite different, in particular as we do not deal with
dynamics on topological subgroups.

{\bf Lemma 7.3.}
{\it Assume that $\psi_n$ is minimal. Then the Haar measure $m$ is the unique
$\psi_n-$invariant  measure on $S_{p^{-l}}(1)$.}

{\noindent \bf Proof. }%
First note that minimality of $\psi_n$ implies that $(n,p)=1$ and hence that $\psi_n$
is an isometry on $S_{p^{-l}}(1)$.
Then, as a consequence of Theorem 27.5 in [23], it follows that 
$\psi_n(U_r(a))=U_r(\psi_n(a))$ for each ball $U_r(a)\subset S_{p^{-l}}(1)$.
Consequently, for every open set $U\neq\emptyset$ we have 
$S_{p^{-l}}(1)=\cup_{N=0}^{\infty}\psi_n^N(U)$. 
It follows for a $\psi_n-$invariant measure $\mu$ that $\mu (U)>0$. 

Moreover we can split $S_{p^{-l}}(1)$ into disjoint balls of radii
$p^{-(l+k)}$, $k\geq 1$, on which $\psi_n$ acts as a permutation. 
In fact, for each $k\geq 1$, $S_{p^{-l}}(1)$ is the union,
\begin{equation}\label{splitting}
S_{p^{-l}}(1)=\cup U_{p^{-(l+k)}}(1+b_lp^l+...+b_{l+k-1}p^{l+k-1}),
\end{equation}
where $b_i\in\{0,1,...,p-1\}$ and $b_l\neq 0$.

We now show that $\psi_n$ is a permutation on the partition
(\ref{splitting}). Recall that  every element of a $p-$adic ball is
the center of that ball, and as
pointed out above $\psi_n(U_r(a))=U_r(\psi_n(a))$. Consequently we
have for all positive integers $k$, $\psi_n^k(a)\in U_r(a)\Rightarrow \psi_n^k(U_r(a))=U_r(\psi_n^k(a))=U_r(a)$ so that
$\psi_n^{Nk}(a)\in U_r(a)$ for every natural number
$N$. Hence, for a minimal $\psi_n$ a point of a ball $B$ of the
partition (\ref{splitting}) must move to another ball in the partition.

Furthermore the minimality of $\psi_n$
shows indeed that $\psi_n$ acts as
a permutation on balls. By invariance of $\mu$ all balls must have the 
same positive measure. As this holds for any $k$, $\mu$ must be the
restriction of Haar measure $m$. 

\vspace{1.5ex}
\noindent
The arguments of the proof of Lemma 7.3 also show that Haar 
measure is always $\psi_n-$invariant. Thus if $\psi_n$ is uniquely ergodic, 
the unique invariant measure must be the Haar measure $m$. Under these 
circumstances it is known [127] that $\psi_n$ must 
be minimal.

{\bf Theorem 7.3.}
{\it The monomial dynamical system $\psi_n:$ $x\mapsto x^n$ on $S_{p^{-l}}(1)$ is 
minimal if and only if it is uniquely ergodic in which case the unique 
invariant measure is the Haar measure.}

\vspace{1.5ex}
\noindent
Let us mention that unique ergodicity yields in particular the ergodicity of 
the unique invariant measure, i.e., the Haar measure $m$, which means that
\begin{equation}
\frac {1}{N}\sum_{i=0}^{N-1}f(x^{n^i})\rightarrow\int f\, dm \; \; \mbox{ for all } \;
x\in S_{p^{-l}}(1),
\end{equation}
and all continuous functions $f:\quad S_{p^{-l}}(1)\rightarrow {\bf R}$.

On the other hand the arguments of the proof of Lemma 7.3, i.e.,
 the fact that $\psi_n$ acts as a permutation on each partition of 
$S_{p^{-l}}(1)$ into disjoint balls if and only if 
$\langle n\rangle =F^\star_{p^2}$,  proves that if $n$ is not a generator of 
$F^\star_{p^2}$ then the system is not ergodic with respect to Haar measure. Consequently, 
if $\psi_n$ is ergodic then $\langle n\rangle =F^\star_{p^2}$ so that the system is 
minimal by Theorem 7.2,
and hence even uniquely ergodic by Theorem 7.3. Since
 unique ergodicity implies ergodicity  one has the following.

{\bf Theorem 7.4.}
{\it The monomial dynamical system $\psi_n:$ $x\mapsto x^n$ on $S_{p^{-l}}(1)$ is 
ergodic with respect to Haar measure if and only if it is uniquely ergodic.}

Even if the monomial dynamical system $\psi_n:$ $x\mapsto x^n$ on
$S_{p^{-l}}(1)$ is ergodic, it never can be mixing, especially not
weak-mixing. This can be seen from the fact that an abstract dynamical
system is weak-mixing if and only if the product of such two systems
is ergodic. If we choose a function $f$ on $S_{p^{-l}}(1)$ and define
a function $F$ on $S_{p^{-l}}(1)\times S_{p^{-l}}(1)$ by
$F(x,y):=f(\log x/\log y)$ (which is well defined as $\log$ does not
vanish on $S_{p^{-l}}(1)$), we obtain a non-constant function
satisfying $F(\psi_n(x),\psi_n(y))=F(x,y)$. This shows, see [127],
that $\psi_n\times\psi_n$ is not ergodic,
and hence $\psi_n$ is not weak-mixing with respect to any invariant
measure, in particular the restriction of Haar measure.

\vspace{1.5ex}
\noindent
Let us consider the ergodicity of a perturbed system
\begin{equation}\label{perturbation}
\psi_q=x^n+q(x),
\end{equation}
for some polynomial $q$ such that $q(x)$ equals to $0 \; \rm{mod}\; p^{l+1}$, 
($\left | q(x)\right |_p<p^{-(l+1)}$). This condition is necessary in
order to guarantee
that the sphere $S_{p^{-l}}(1)$ is invariant. For such a system to be ergodic
 it is necessary that $n$ is a generator of $F^\star_{p^2}$. This follows from the
 fact that for each $x=1+a_lp^l+...$  on $S_{p^{-l}}(1)$ 
(so that $a_l\neq 0$) the condition on $q$ gives 
\begin{equation}
\psi_q^N(x) \;\;\mbox{ equals to}\; 1+n^Na_l  \;\rm{mod}\; p^{l+1}.
\end{equation}
Now $\psi_q$ acts as a permutation on the $p-1$ balls of radius 
$p^{-(l+1)}$ if and only if $\langle n\rangle =F^\star_{p^2}$. Consequently, a 
perturbation (\ref{perturbation}) cannot make a nonergodic system ergodic.

\medskip

{\bf Appendix 1:  Newton's Method (Hensel's Lemma)}

Here we present a $p$-adic analogue of the Newton procedure to find 
the roots of polynomial equations, see e.g. [2]:

{\bf Theorem.} {\it Let $F(x), \; x\in {\bf Z}_p ,$ be a polynomial with coefficients
$F_i\in {\bf Z}_p.$ Let there exists $\gamma\in {\bf Z}_p$ such that
$$
F(\gamma) =0 \; (\rm{mod}\; p^{2\delta +1} ) \mbox{ and } 
F^\prime (\gamma ) =0 \; (\rm{mod}  \; p^\delta),  \; 
F^\prime (\gamma )\not= 0 \; (\rm{mod}  \; p^{\delta+1}) ,
$$
where $\delta $ is a natural number. Then there exists a $p$-adic integer
$\alpha $ such that
$$
F(\alpha) = 0 \mbox{ and } \alpha =\gamma \; (\rm{mod}   \; p^{\delta+1}).
$$}

{\bf Corollary} (Hensel Lemma). {\it Let $p(x)$ be a polynomial with $p$-adic integer
coefficients and let there exist $\gamma\in {\bf Z}_p$ such that:
$$
F(\gamma)=0 \; (\rm{mod}  \; p),\; \; \;  F^\prime (\gamma)\not= 0 \;
(\rm{mod}  \; p).
$$
Then there exists $\alpha\in {\bf Z}_p$ such that 
$$
F(\alpha ) =0 \;\; \; \mbox{and} \; \; \; \alpha =\gamma \; (\rm{mod}  \; p).
$$ }

{\bf Appendix 2: Computer Calculations for Fuzzy Cycles}

The following results were obtained with the aid of the complex 
of  $p$-adic programs, {\it $p$-ADIC}, which was created by De Smedt, see e.g. [92],
using the standard software package MATHEMATICA.

{\bf Example.} Consider the function $\psi_3(x)= x^3$ in ${\bf Q}_5$. 
Then we found among others the following fuzzy cycles. Cycles of length 2 :
$$U_{1 \over 5}(2) - U_{1 \over 5}(3);\;\;
U_{1 \over 25}(7) - U_{1 \over 25}(18);\;\;
U_{1 \over 125}(57) - U_{1 \over 125}(68).
$$
Cycles of length 4 :
$$
\begin{array}{rl}
&
U_{1 \over 25}(6) - U_{1 \over 25}(16) - U_{1 \over 25}(21) - U_{1
\over 25}(11);
\\ & \\
&
U_{1 \over 25}(2) - U_{1 \over 25}(8) - U_{1 \over 25}(12) - U_{1 \over
25}(3);
\\ & \\
&
U_{1 \over 25}(22) - U_{1 \over 25}(23) - U_{1 \over 25}(17) - U_{1
\over 25}(13);
\\ & \\
&
U_{1 \over 25}(9) - U_{1 \over 25}(4) - U_{1 \over 25}(14) - U_{1
\over 25}(19);
\\ & \\
&
U_{1 \over 125}(7) - U_{1 \over 125}(93) - U_{1 \over 125}(107) - U_{1
\over 125}(43);
\\ & \\
&
U_{1 \over 125}(26) - U_{1 \over 125}(76) - U_{1 \over 125}(101) -
U_{1 \over 125}(51);
\\ & \\
&
U_{1 \over 125}(18) - U_{1 \over 125}(82) - U_{1 \over 125}(112) -
U_{1 \over 125}(32);
\\ & \\
&
U_{1 \over 125}(24) - U_{1 \over 125}(74) - U_{1 \over 125}(99) - U_{1 \over 125}(49).
\\
\end{array}
$$
Cycles of length 20 :
$$
\begin{array}{rl}
&
U_{1 \over 125}(6) - U_{1 \over 125}(91) - U_{1 \over 125}(71) - U_{1
\over 125}(36) -
U_{1 \over 125}(31)
\\ & \\
&
- U_{1 \over 125}(41) - U_{1 \over 125}(46) -
U_{1 \over 125}(86) - U_{1 \over 125}(56) - U_{1 \over 125}(116)
\\ & \\
&
-U_{1 \over 125}(21) -
U_{1 \over 125}(11) - U_{1 \over 125}(81) - U_{1 \over 125}(66) 
\\ & \\
&
- U_{1 \over 125}(121) - U_{1 \over 125}(61) - U_{1 \over 125}(106)
\\ & \\
&
-U_{1 \over 125}(16) -
U_{1 \over 125}(96) - U_{1 \over 125}(111)
\\
\end{array}
$$

$p$-ADIC gives the possibility of studying more complicated dynamical systems.
However, we cannot find exact cycles with the aid of a computer.
As a consequence
of a finite precision we can find only fuzzy cycles. Therefore we shall
study fuzzy cycles and their behaviour. We define {\it fuzzy attractors} and
{\it fuzzy Siegel disks} by direct generalization of the corresponding definitions for point cycles. 
Let $f(x)= x^2 + x.$ The following fuzzy cycles were found in the case where
$p$ is a prime less than 100.

Cycles of length 2 for $p = 5, 13, 17, 29, 37, 41, 53, 61, 73, 89, 97.$
Moreover, we have proved the following general statement.

{\bf Proposition.} {\it Let $p = 1 (\rm{mod}\; 4).$ Then the dynamical system
$f(x)= x^2 + x$ has fuzzy cycles of length {\rm{2}}. In case $p = 5$ these
are fuzzy cyclic attractors, and in the other case these are Siegel disks.}

Cycles of length 3 for $p = 11, 41, 43, 59, 67, 89$ (twice), 97.  In the case $p =
89$ one of the fuzzy cycles is an attractor, all others are Siegel disks.  

Cycles of length 4 for $p = 19, 43, 47, 71$ (all Siegel disks).

Cycles of length 5 for $p = 23, 41, 71, 73$ (all Siegel disks).

Cycles of length 6 for $p = 47, 83, 89$ (all Siegel disks).

Cycles of length 7 for $p = 29, 53, 59, 67$ (cyclic
attractors in the case $p = 29).$

Cycles of length 8 for $p = 61$ (all Siegel disks).

Cycles of length 9 for $p = 31$ (all Siegel disks).

Remark that for some primes we have fuzzy cycles of different 
lengths. There are fuzzy cycles of length 2, 3 and 6, for example, for $p = 89.$
There are fuzzy cycles of length 2, 3 and 5 and for $p = 41.$ 

Some of these cycles (we think all of them, but we have not
proved it) contain subcycles.  For example, in the case $p = 11$
we have the cycle of length 3:
$U_{1/11}(2) - U_{1/11}(6) - U_{1/11}(9),$
which contains subcycles of length 15:
$$
U_{1/121}(112) - U_{1/121}(72) - U_{1/121}(53) - U_{1/121}(79) - U_{1/121}(28) 
$$
$$
- U_{1/121}(86) -
U_{1/121}(101) - U_{1/121}(17) - U_{1/121}(64) - U_{1/121}(46)
$$
$$
- U_{1/121}(105) - 
U_{1/121}(119) - U_{1/121}(2) - U_{1/121}(6) - U_{1/121}(42)
$$
and
$$
U_{1/121}(35) - U_{1/121}(50) - U_{1/121}(9) - U_{1/121}(90) - U_{1/121}(83)
$$
$$
-U_{1/121}(75)-
U_{1/121}(13) - U_{1/121}(61) - U_{1/121}(31) - U_{1/121}(24)
$$
$$
- U_{1/121}(116) - U_{1/121}(20) 
- U_{1/121}(57) - U_{1/121}(39) - U_{1/121}(108)
$$
and so on.

In the case $p = 13$ we have the cycle of length 2:
$U_{1/13}(4) - U_{1/13}(7)$
which contains, amongst others, the subcycle of length 8:
$$
U_{1/169}(4) - U_{1/169}(20) - U_{1/169}(82) - U_{1/169}(36)
$$
$$
- U_{1/169}(134) - U_{1/169}(7) - U_{1/169}(56) - U_{1/169}(150)
$$
which contains, amongst others, subcycles of length 104.

One of the problems which arise in our computer investigations of $p$-adic dynamical systems
is that we cannot  propose a reasonable way of creating
$p$-adic pictures which can illustrate our numerical results. However,
this
is a general problem of the $p$-adic framework because the human brain
can
understand only pictures in real space.

The author would like to thank S. Albeverio, I. Volovich, 
V. Vladimirov, I. Aref'eva, G. Parisi, B. Dragovich, S. Kozyrev for discussions on $p$-adic physics, 
L. Arnold and M. Gundlach --  $p$-adic random dynamical systems, D. Amit, B. Hiley, P. Kloeden and B. Tirozzi 
-- $p$-adic cognitive models, W. Schikhof, A. Escassut, M. Endo, L. van Hamme -- $p$-adic analysis, 
R. Benedetto, L. Hsia and F. Vivaldi -- $p$-adic dynamical systems.

\medskip

{\bf References}

1. K. Hensel, Untersuchung der Fundamentalgleichung einer Gattung f\"ur
eine reelle Primzahl als Modul und Bestimmung der Theiler ihrer
Discriminante, {\it J. Reine Angew Math.}, 1894, vol. 113, pp. 61--83.

2.  Z. I. Borevich and I. R. Shafarevich, {\it Number Theory},
New York: Academic Press, 1966.

3.  J. K\"urschak, \"Uber Limesbildung und allgemeine K\"opertheorie,
{\it J. Reine Angew. Math.},  1913, vol. 142, pp. 211-253.

4. A. Ostrowski, {\it J. Reine Angew. Math.}, 1917, vol. 147, pp. 191-- 204;
{\it  Acta Math.},  1917, 41, pp. 271--284.

5. L. S. Pontryagin, {\it Ann. of Math.},  1932,(2) vol. 33,  pp. 163--174.

6. N. Jacobson, Totally disconnected locally compact rings, {\it Amer. J. Math.},
1936, 58, pp. 433-449.

7. W. Krull, Allgemeine Bewertungstheorie, {\it J. Reine Angew. Math.}, 1932,
vol. 167, pp. 160-196.

8. S. MacLane, A construction for absolute values in polynomial rings,
{\it Trans. Am. Math. Soc.}, 1936, vol.  40, pp. 363-395.

9. I. Kaplansky, Topological rings, {\it Bull. A. Math. Soc.}, 1948, vol. 54, pp. 809-826.

10. I. R.  Shafarevich, On the normalizability of topological rings,
{\it DAN SSSR}, 1943, vol. 40, pp. 133-135.

11. D. Zelinsky, Topological characterization of fields with valuations,
{\it Duke Math.}, 1948, vol. 15, pp. 595-622.

12. H.-J. Kowalsky, Beitr\"age zur topologischen Algebra,
{\it Math. Nachr.}, 1954, vol. 11, pp. 143-185.

13.  K. Mahler, {\it $p$-adic Numbers and their Functions},
Cambridge tracts in math., vol. 76. Cambridge: Cambridge Univ. Press, 1980.

14.  M. Krasner, Nombres semi-reels et espaces ultrametriques, {\it C. R. Acad. Sci. Paris,} 
1944, vol. 219, pp. 433-435.

15.  M. Krasner, Prolongement analytique uniforme et multiforme dans 
les corps valu\'es complets: pr\'eservation de l'analycit\'e par la convergence uniforme, Th\'eor\`eme
de Mittag-Leffler g\'en\'eralis\'e pour les \'el\'ementes analytiques,
{\it C.N.R.S. Paris, A,} 1957, vol. 244, pp. 2570-2573.

16. F. Hausdorff, Erweiterung einer Homeomorphie,
{\it Fund. Math.}, 1930, vol. 16, pp. 353-360; 
Unber innere Abbildungen, {\it Ibid}, 1934, vol. 23, pp. 279-299.

17. E. Motzkin, Ph. Robba, Prolongement analytique en
analyse $p$-adique,´{\it  S\'eminarie de theorie des nombres,} 1968-1969,
Fac. Sc. de Bordeaux.

18.  Ph. Robba, Fonctions analytiques sur les corps valu\'es 
ultra- m\'etriques complets. Prolongement analytique et alg\`ebres 
de Banach ultram\'etriques, {\it Ast\' erisque}, 1973, no. 10, 
pp. 109--220. 

19. Y. Amice, Les nombres $p$-adiques, P. U. F., 1975.

20. M. Lazard, Les z\'eros des fonctions analytiques sur un corps valu\'e
complet, {\it IHES, Publ. Math.}, 1962, no. 14, pp. 47-75.

21.  A. Monna, {\it Analyse non-Archim\'edienne}, New York: Springer-Verlag,  1970.

22. T. A. Springer, Quadratic forms over fields with a
discrete valuation, 1, {\it Proc. Kon. Ned. Akad. v. Wetensch.}, 1955, vol. 58, pp. 352--362.

23.  W.H. Schikhof, \emph{Ultrametric calculus}, Cambridge: Cambridge
University press, 1984.

24. A. M. Robert, {\it A course in $p$-adic analysis}, New York: Springer-Verlag, 2000.

25. A. Escassut, {\it Analytic elements in $p$-adic analysis},
Singapore: World Scientic, 1995.

26. P.-C. Hu, C.-C. Yang, {\it Meromorphic functions over Non-Archimedean fields},
Dordrecht: Kluwer Academic Publishers, 2000.

27. J. Tate, Rigid analytic spaces, {\it Invent. Math.}, 1971, vol.12, 257-289.

28. S. Bosch, U. G\"untzer , R. Remmert, {\it  Non-Archimedean
analysis}, Berlin-Heidelberg-New York: Springer-Verlag, 1984.

29.  S. Bosch , Orthonormalbasen in der nichtarchimedischen 
Funktiontheorie, {\it Manuskripta Math.}, 1969, vol. 1, pp. 35-57.

30.  S. Bosch , A rigid analytic version of M.Artin`s theorem on analytic
equations, {\it Math. Ann.}, 1981, vol. 255, pp. 395-404.

31. L. Gerritzen , Erweiterungsendliche Ringe in der
nichtarchimedischen Funktionentheorie,
{\it Invent. Math.}, 1966, vol. 2, pp. 178-190.

32.  L. Gerritzen, H. Grauert, Die Azyklizit\"at der affinoiden
\"Uberdeckungen. Global Analysis, {\it Papers in Honor of K. Kodaira.}
Princenton: Princenton Univ.Press, pp. 159-184, 1969.

33.  L. Gerritzen, U. G\"untzer, \"Uber Restklassennormen auf affinoiden
Algebren, {\it Invent. Math.}, 1967, vol. 3, pp. 71-74.

34. H. Grauert, R. Remmert , Nichtarchimedische Funktiontheorie, {\it Weierstrass-
Festschrift, Wissenschaftl. Abh. Arbeitsgemeinschaft f\"ur Forschung des 
Landes Nordrhein-Westfalen}, 1966, vol. 33, pp. 393-476.

35.  H. Grauert, R. Remmert , \"Uber die Methode der diskret bewerteten Ringe
in der nicht-archimedischen Analysis,  {\it Invent. Math.}, 1966, vol. 2, pp. 87-133.

36. Ya. Morita, On the induced $h$-structure on an open subset of the rigid analytic 
space $P^1(k),$ {\it Math. Annalen,} 1979, vol. 242, pp. 47-58.

37. Ya. Morita, A $p$-adic theory of hyperfunctions, {\it Publ. RIMS}, 1981, no. 1, 
pp.1-24.

38. T. Kubota, H.-W. Leopoldt, Eine $p$-adische Theorie der Zetawerte, 1. 
Einf\"uhrung der $p$-adischen Dirichletschen $L$-Funktionen,
{\it J. Reine Angew. Math.}, 1964, vol. 214/215, pp. 328-339.

39. H.-W. Leopoldt, Eine $p$-adische Theorie der Zetawerte, 2. 
Die $p$-adische $\Gamma$-Transformation. Collection of articles dedicated to Helmut 
Hasse on his 75th birthday, {\it J. Reine Angew. Math.}, 1975, vol. 274/275, pp. 224-239.

40. T. Kubota, Local relation of Gauss sums, {\it Acta Arith.}, 1960/61, vol. 6, 285-294.

41. K. Iwasawa, {\it Lectures on $p$-adic $L$-function,}
Princeton: Princeton Univ. Press, 1972.

42. Ya. Morita, On the radius of convergence of the $p$-adic $L$-function,
{\it Nagoya Math. J.}, 1979, vol. 75, pp. 177-193.

43. Ya. Morita, A $p$-adic analogue of the $\Gamma$-function,
{\it J. Fac. Sc. Univ. Tokyo, Sect. IA, Math.,} 1975, vol. 22, no.2, pp. 255-266.

44.  I. M. Gelfand, M. I. Graev and I. I.  Pjatetskii-Shapiro, 
{\it Representation theory and automorphic functions,} London: Saunders, 1966.

45.  I. M. Gelfand and M. I. Graev, Irreducible unitary representations
of the group of matrices of the second order with elements from a locally compact field,
{\it Dokl. Acad. Nauk SSSR,} 1963, vol. 149, pp. 499-501.

46.  I. M. Gelfand and M. I. Graev, Representations
of the group of matrices of the second order with elements from a locally compact field,
and special functions on locally compact field,
{\it Uspehi Mat. Nauk,}   1963, vol. 18, no. 4, pp. 29-99.

47. I. M. Gelfand, M. I. Graev and I. I.  Pjatetskii-Shapiro, Representations
of adele groups, {\it Dokl. Acad. Nauk SSSR,} 1964, vol. 156, pp. 487-490.

48.  I. M. Gelfand and M. I. Graev, The structure of the ring of finite functions on the group of second-order
unimodular matrices with elements belonging to a disconnected locally compact field,
{\it Soviet Math. Dokl.,} 1963, vol. 4, pp. 1697-1700.

49. M. I. Graev and R. I. Prohorova, Homogeneoius generalized functions in a vector space over a
local nonarchimedean field that are connected with a quadratic form. {\it Functional Anal. and
Appl.,}  1972, vol. 6, no. 3, pp. 70-71.

50. R. I. Prohorova, Action of the group $lo(n)$ in a vector space over a
local nonarchimedean field, {\it Functional Analysis}, 1977, no. 9, pp. 129-137.

51. A. A. Kirillov and R. R. Sundcheleev, Algebra of measures on the group of affine transformations 
of the $p$-adic interval,  {\it Dokl Acad. Nauk UzbSSR,} 1975, vol. 2, pp. 3-4.

52.  A. D. Gvishiani, Representations of the group of local translations of the space $k^m$ 
where $k$ is a non-Archimedean field, {\it Functional Anal. and
Appl.,}   1979, vol. 13, no. 3, pp. 73-74.

53. P. M. Gudivok, Modular and integer $p$-adic representations of a direct product of groups,
{\it Ukr. Mat. Z.}, 1977, vol. 29, no. 5, pp. 580-588.

54. A. V. Zelevinskii, Classification of irreducible noncuspidal  representations 
of the group ${\rm GL}(n)$ over a $p$-adic field, {\it Functional Anal. and
Appl.,} 1977, vol. 11, no. 1, pp.67-68.

55. A. V. Trusov, The principal series  of representations of a group of $p$-adic quaternions in spaces over
non-Archimedean fields, {\it Uspehi Mat. Nauk,} 1982, vol. 37, no. 4 (226), pp. 181-182.

56. A. V. Trusov, Representations of the groups ${\rm GL} (2, Z_p)$ and  ${\rm GL} (2, Q_p)$
in the spaces over non-Archimedean fields, {\it Vestnik Moskov. Univ., Ser. 1, Mat. Mekh.}, 
1981, no. 1, pp. 55-59.

57. A. Monna  and  F. van der Blij,  Models of space and time in elementary physics,
{\it J. Math. Anal. and Appl.}, 1968, vol. 22, pp. 537--545.

58. E. Beltrametti  and G. Cassinelli, Quantum mechanics and $p$-adic
numbers, {\it Found. Phys.}, 1972, vol. 2, pp. 1-7.

59.  Yu. Manin, New dimensions in geometry, {\it Lect. Notes in Math.},
1985, vol. 1111, pp. 59--101.

60.  I. V. Volovich,  $p$-adic string, {\it Class. Quant. Grav.},
1987, vol. 4, pp. 83--87.

61. P. G. O. Freund  and E. Witten, Adelic string amplitudes,
{\it Phys. Lett. B}, 1987, vol. 199, pp. 191-195.
                    
62.  G. Parisi, $p$-adic functional integral, 
 {\it Mod. Phys. Lett.} A, 1988, vol. 4,  pp. 369-374.

63. I. Ya. Aref'eva,  B. Dragovich ,  P. H. Frampton,  I. V. Volovich
The wave function of the Universe and $p$-adic gravity, {\it Int. J. of 
Modern Phys.} A, 1991, vol. 6, no 24, pp. 4341--4358.

64. L. Brekke, P. G. O. Freund, E. Melzer and M. Olson,
Adelic $N$-point amplitudes, {\it Phys. Lett.} B, 1989, vol. 216, pp. 123-126.

65.  V. S. Vladimirov,  I.  V. Volovich, and  E. I. Zelenov, 
{\it $p$-adic analysis and  Mathematical Physics}, Singapore: World Scientific Publ., 
1994.

66.   V. S.  Vladimirov  and I.  V. Volovich,  
$p$-adic quantum mechanics, {\it Commun. Math. Phys.}, 1989, vol.  123, pp. 659--676.

67. P. Ruelle, E. Thiran, D. Verstegen and J. Weyers, Adelic string and superstring
amplitudes. {\it Mod. Phys. Lett.} A, 1989, vol. 4, pp. 1745-1753.

68.  A. Yu. Khrennikov,  $p$-adic quantum mechanics with $p$-adic
valued functions,     {\it J. Math. Phys.}, 1991, vol. 32, no. 4, pp. 932--937.

69.  A. Yu. Khrennikov, \emph{$p$-adic Valued Distributions in
Mathematical Physics}, Dordrecht: Kluwer Academic Publishers, 1994.

70.  S. Albeverio S.,  A.Yu. Khrennikov, Representation of the Weyl group
in spaces of square integrable functions with respect to $p$-adic valued 
Gaussian distributions,  {\it  J. of Phys. A: Math. Gen.},  1996, vol. 29, pp. 5515--5527.

71.  G. Parisi, N. Sourlas, $p$-adic numbers and replica smmetry breaking.
{\it The European Physical J.}, B, 2000, vol. 14, pp. 535-542.

72. V. A. Avetisov, A. H. Bikulov, S. V. Kozyrev,
Application of $p$-adic analysis to models of breaking 
of replica symmetry, {\it J. Phys. A: Math. Gen.,} 1999, vol.  32, pp. 8785-8791.

73. B. Dwork, On $p$-adic differential equations, 1. The Frobenius structure of differential equations,
{\it Bull. Soc. Math. France,} 1974, no. 39/40, pp. 27--37.

74. B. Dwork, P. Robba, On ordinary linear $p$-adic differential equations witha algebraic
functional coefficients, {\it  Trans. Amer. Math. Soc.,} 1977, vol. 231, no. 1, pp. 1-46.

75.   B. Dwork, {\it Lectures on $p$-adic differential equantions}, 
 Berlin--Heidelberg--New York: Springer-Verlag, 1982.

76.  B. Dwork, G. Gerotto, F. J. Sullivan,  An introduction to $G$-functions,
 {\it Annals of Math. Studies}, Princeton:  Princeton Univ. Press, 1994.

77.  G. Christol, Modules differentiels et equations differentielles
$p$--adiques. {\it Queen`s papers in pure and applied mathematics,}
vol.  66, Kingston: Queen`s Univ. Press, Ontario, 1983. 

78.   A. Yu. Khrennikov, Mathematical
methods of the non-Ar\-chi\-me\-de\-an physics,
{\it Uspekhi Mat. Nauk}, 1990, vol. 45, no. 4, pp.7 9--110.

79.  A. Yu. Khrennikov, Fundamental solutions over the field of
$p$-adic numbers, {\it Algebra and Analysis(Leningrad Math.J.)},  1992, vol. 4, no. 3, pp. 248--266.

80.  A. Yu. Khrennikov,   Trotter's  formula  for   heat
conduction equations and for Schr\"odinger's equation
on non-Archimedean superspace,    {\it Siberian  Math. J.}, 1991, vol. 32,
no. 5, pp. 155--165.

81.   A. Escassut,  A. Yu. Khrennikov, Nonlinear differential equations over 
the field of complex $p$-adic numbers.  In: $p$-adic functional analysis.
Editors: Schikhof W., Perez-Garcia C., Kakol J.; 
{\it Lecture Notes in Pure and Applied Mathematics,} 1997, vol. 192, pp. 143-151.

82.   N. De Grande - De Kimpe,  A.Yu. Khrennikov, The non-Archimedean
Laplace transform, {\it Bull. Belgian Math. Soc.,} 1996,  vol.  3, pp. 225-237.

83.  N. De Grande-De Kimpe, A. Yu. Khrennikov, L. Van Hamme, The Fourier transform for
$p$-adic smooth distributions, {\it Lecture Notes in Pure and Applied Mathematics,} New York: Dekker,
1999, vol. 207, pp. 97-112.

84. E. Thiran,  D. Verstegen, J. Weyers, $p-$adic dynamics,
\emph{J. Stat. Phys.}, 1989, vol. 54, pp. 893-913.

85. D. K. Arrowsmith,  \&  F. Vivaldi, Some $p-$adic
representations of the Smale horseshoe,
\emph{Phys. Lett.} A, 1993, vol. 176, pp. 292-294.

86. D. K. Arrowsmith, \& F. Vivaldi, Geometry of $p-$adic
Siegel discs,  \emph{Physica} D,  1994, vol. 74, 222-236.

87. J. Lubin, Non-Archimedean dynamical systems, \emph{Compositio Mathematica}, 1994, 
vol. 94,  pp. 321-346.

88. T. Pezda, Polynomial cycles in certain local domains,
\emph{Acta Arithmetica}, vol. LXVI, 1994, pp. 11-22.

89. A. Yu. Khrennikov,
{\it A $p$-adic behaviour of the standard dynamical systems,}
Preprint Ruhr University Bochum, SFB-237, 1995,  no. 290.

90. L. Hsia, A weak N\'eron model with applications to $p-$adic dynamical systems.
\emph{Compositio Mathematica}, 1996, vol. 100,  pp. 227-304.

91.  A. Yu.  Khrennikov, {\it Non-Archimedean analysis: quantum
paradoxes, dynamical systems and biological models,}
Dordreht: Kluwer Academic Publishers, 1997.

92. S. De Smedt, A. Yu. Khrennikov, Dynamical systems and theory of numbers. 
{\it Comment. Math. Univ. St. Pauli,} 1997, vol. 46, no.2, pp. 117-132.

93.  R. Benedetto, Fatou components in $p-$adic dynamics, PHD thesis,
Brown University,  1998.

94. R. Benedetto, Hyperbolic maps in $p-$adic dynamics. 
{\it Ergodic Theory and Dynamical Systems,} 2001, vol. 21, pp. 1-11.

95.  R. Benedetto, Reduction, dynamics, and Julia sets of rational functions.
{\it J. of Number Theory,}  2001, vol. 86, pp. 175-195.

96.  A. Yu. Khrennikov, Small denominators in complex $p$-adic dynamics,
{\it Indag. Mathem.}, 2001, vol. 12 (2), pp. 177-189.

97.  K.-O. Lindahl, {\it Dynamical Systems in $p$-adic geometry.}
Licentiate-thesis, Reports from MSI, no. 01076, V\"axj\"o: V\"axj\"o University Press, 2001.

98.   M. Gundlach, A. Khrennikov, K.-O. Lindahl, Topological transitivity for p-adic
dynamical systems. {\it p-adic functional analysis.} Ed. A. K. Katsaras, W.H. Schikhof, L. Van Hamme,
{\it Lecture notes in pure and apllied mathematics,} 2001, vol. 222, pp. 127-132.

99.   W. Parry, Z. Coelho,  Ergodicity of $p$-adic multiplication and the distribution of
Fibonacci numbers. {\it  AMS Contemporary Mathematics,} to be published.

100. S. Albeverio, M. Gundlach,  A. Yu. Khrennikov, K.-O. Lindahl, 
On Markovian behaviour of $p$-adic random dynamical systems, {\it Russian J. of Math. Phys.}, 
2001, vol. 8 (2), pp. 135-152.

101. A.Yu. Khrennikov, M. Nilsson, On the number of cycles for $p$-adic dynamical systems, {\it J. Number
Theory}, 2001, vol. 90, pp. 255-264.

102.  M. Nilsson, Cycles of monomial dynamical systems over the field of 
$p-$adic numbers. Reports from V\"axj\"o University, no. 20,  1999.

103. R. Nyqvist, Some dynamical systems in finite field extensions of the $p$-adic
numbers.  {\it $p$-adic functional analysis.} Ed. A. K. Katsaras, W.H. Schikhof, L. Van Hamme,
{\it Lecture notes in pure and applied mathematics,} 2001, vol. 222, pp. 243-254.

104. P.-A. Svensson, {\it Finite extensions of local fields.} Licentiate-thesis, Reports from MSI, 
no. 01062, V\"axj\"o: V\"axj\"o University Press,  2001.

105. A. Yu. Khrennikov, M. Nilsson, N. Mainetti, Non-Archimedean dynamics.  In {\it P-adic numbers in 
number theory, analytic geometry and functional analysis.} Collection of papers in honour N. De Grande-De Kimpe
and L. Van Hamme. Ed. S. Caenepeel. {\it Bull. Belgian Math. Society}, 2002, december, pp. 141-147.

106. A. Yu. Khrennikov, M. Nilsson, Behaviour of Hensel perturbations of $p$-adic 
monomial dynamical systems, {\it Analysis Mathematica,} 2003, vol. 29, pp. 107-133.

107. A. Yu. Khrennikov, M. Nilsson, R. Nyqvist, The asymptotic number of periodic points of discrete
polynomial $p$-adic dynamical systems. {\it Contemporary
Math.}, 2003, vol.  319, pp. 159-166.

108. A. Yu. Khrennikov, S. Ludkovsky, Stochastic processes in non-Archimedean spaces with values in non-Archimedean 
fields, {\it Markov Processes and Related Fields}, 2003, vol. 9, no. 1, pp. 131-162. 

109. C. Woodcock, N. Smart, p-adic chaos and random number generation,
{\it Experimental Math.}, 1998, vol. 7, 333-342.

110. J. Bryk, C. E. Silva, $p$-adic measurable dynamical systems of symple polynomials,
{\it Amer. Math. Monthly}, 2004, to be published.

111. Proceedings of Workshop: "Dynamical systems: from probability to number theory-2", Vaxjo, November-2002,
ed: A. Yu. Khrennikov, Ser. Math. Modelling in Phys., Engin., and Cogn. Sc., vol. 6,
V\"axj\"o Univ. Press, 2003.

112.  L. Arnold, {\it   Random dynamical systems,} Berlin-New York-Heidelberg: Springer-Verlag, 1998.

113.  A. Yu. Khrennikov, Human subconscious as the p-adic dynamical 
system, {\it J. of Theor. Biology,} 1998, vol. 193, pp. 179-196.

114. S. Albeverio, A. Yu. Khrennikov, P. Kloeden, 
    Memory retrieval as a $p$-adic dynamical system, {\it Biosystems,} 1999,  vol. 49,
    pp. 105-115.

115.  D. Dubischar, V. M. Gundlach, O. Steinkamp,
A. Yu.  Khrennikov,  Attractors of random dynamical systems over
$p$-adic numbers and a model of noisy cognitive processes, {\it Physica} D,
1999, vol. 130, pp. 1-12.

116. S. Albeverio , A. Yu. Khrennikov,
B. Tirozzi, $p$-adic neural networks. {\it Mathematical models and methods in 
applied sciences,} 1999, vol. 9, no. 9, pp. 1417-1437.

117.  A. Yu. Khrennikov, $p$-adic information spaces, infinitely small probabilities 
and anomalous phenomena,
{\it J. of Scientific Exploration,} 1999, {\bf 4}, no.13, pp. 665-680.

118. A. Yu. Khrennikov, $p$-adic discrete dynamical systems and collective behaviour of 
information states in cognitive models,
{\it Discrete Dynamics in Nature and Society,} 2000, vol. 5, pp. 59-69.

119. A. Yu. Khrennikov, Informational interpretation of $p$-adic physics,
{\it Dokl. Akad. Nauk, } 2000, vol. 373, no. 2, vol. 174-177.

120.  A. Yu. Khrennikov, Classical and quantum mechanics on $p$-adic trees of ideas.
{\it BioSystems,} 2000, vol. 56, pp. 95-120.

121. A. Yu. Khrennikov, $p$-adic model of hierarchical intelligence, {\it Dokl. Akad. Nauk.}, 2003,
vol. 388, no. 6, pp. 1-4.

122. A. Yu. Khrennikov, Quantum-like formalism for cognitive measurements, {\it Biosystems,}
2003, vol. 70, pp. 211-233.

123. J. Benois-Pineau, A. Yu. Khrennikov and N. V. Kotovich, Segmentation of images in $p$-adic 
and Euclidean metrics, {\it Doklady Mathematics}, 2001, vol, 64, no. 3, pp. 450-455; translation
from {\it Dokl. Akad. Nauk.},  2001, vol. 381, no. 5, pp. 604-609.

124. A. Yu. Khrennikov and N. V. Kotovich, $m$-adic coordinate representation of 
images, {\it Dokl. Akad. Nauk.}, 2002, vol. 387, no. 2, pp. 115-119.

125.  J. Hadamard, Sur la distribution des zéros de la
	fonction $\zeta(s)$ et ses conséquences arithmétiques,
  {\it Bull. Soc. Math. France,} 1896, vol. 24, pp. 199-220.

126.  T. M. Apostol, {\it Introduction to analytic number
theory}, Berlin--Heidelberg--New York: Springer-Verlag, 1976.
  
127.  P. Walters, {\it  An introduction to ergodic theory,}
Berlin--Heidelberg--New York: Springer-Verlag,  1982.

\end{document}